\documentclass[iop]{emulateapj}

\pdfoutput=1

\usepackage{bm}
\usepackage{subfigure}
\usepackage{amsmath}
\usepackage{multirow}
\usepackage[colorlinks,citecolor=blue,linkcolor=blue,urlcolor=blue]{hyperref}

\usepackage{etoolbox}

\makeatletter
\patchcmd{\NAT@citex}
  {\@citea\NAT@hyper@{%
     \NAT@nmfmt{\NAT@nm}%
     \hyper@natlinkbreak{\NAT@aysep\NAT@spacechar}{\@citeb\@extra@b@citeb}%
     \NAT@date}}
  {\@citea\NAT@nmfmt{\NAT@nm}%
   \NAT@aysep\NAT@spacechar\NAT@hyper@{\NAT@date}}{}{}
\patchcmd{\NAT@citex}
  {\@citea\NAT@hyper@{%
     \NAT@nmfmt{\NAT@nm}%
     \hyper@natlinkbreak{\NAT@spacechar\NAT@@open\if*#1*\else#1\NAT@spacechar\fi}%
       {\@citeb\@extra@b@citeb}%
     \NAT@date}}
  {\@citea\NAT@nmfmt{\NAT@nm}%
   \NAT@spacechar\NAT@@open\if*#1*\else#1\NAT@spacechar\fi\NAT@hyper@{\NAT@date}}
  {}{}
\makeatother

\shorttitle{Combined Maximum-Likelihood Analysis of IceCube High-Energy Data}
\shortauthors{M. G. Aartsen et al.}

\begin{document}

\journalinfo{Preprint accepted for publication in The Astrophysical Journal}
\submitted{}

\title{A combined maximum-likelihood analysis of the high-energy astrophysical neutrino flux measured with IceCube}

\author{
M.~G.~Aartsen\altaffilmark{1},
K.~Abraham\altaffilmark{2},
M.~Ackermann\altaffilmark{3},
J.~Adams\altaffilmark{4},
J.~A.~Aguilar\altaffilmark{5},
M.~Ahlers\altaffilmark{6},
M.~Ahrens\altaffilmark{7},
D.~Altmann\altaffilmark{8},
T.~Anderson\altaffilmark{9},
M.~Archinger\altaffilmark{10}
% ...
(The IceCube Collaboration)
}
\affil{
$^{1}$ Department of Physics, University of Adelaide, Adelaide, 5005, Australia\\
$^{2}$ Technische Universit\"at M\"unchen, D-85748 Garching, Germany\\
$^{3}$ DESY, D-15735 Zeuthen, Germany\\
$^{4}$ Dept.~of Physics and Astronomy, University of Canterbury, Private Bag 4800, Christchurch, New Zealand\\
$^{5}$ Universit\'e Libre de Bruxelles, Science Faculty CP230, B-1050 Brussels, Belgium\\
$^{6}$ Dept.~of Physics and Wisconsin IceCube Particle Astrophysics Center, University of Wisconsin, Madison, WI 53706, USA\\
$^{7}$ Oskar Klein Centre and Dept.~of Physics, Stockholm University, SE-10691 Stockholm, Sweden\\
$^{8}$ Erlangen Centre for Astroparticle Physics, Friedrich-Alexander-Universit\"at Erlangen-N\"urnberg, D-91058 Erlangen, Germany\\
$^{9}$ Dept.~of Physics, Pennsylvania State University, University Park, PA 16802, USA\\
$^{10}$ Institute of Physics, University of Mainz, Staudinger Weg 7, D-55099 Mainz, Germany\\
% ...
}
\altaffiltext{$\ast$}{Corresponding author (\href{mailto:lars.mohrmann@desy.de}{lars.mohrmann@desy.de}).}

\begin{abstract}
Evidence for an extraterrestrial flux of high-energy neutrinos has now been found in multiple searches with the IceCube detector.
The first solid evidence was provided by a search for neutrino events with deposited energies $\gtrsim30$~TeV and interaction vertices inside the instrumented volume.
Recent analyses suggest that the extraterrestrial flux extends to lower energies and is also visible with throughgoing, $\nu_\mu$-induced tracks from the Northern hemisphere.
Here, we combine the results from six different IceCube searches for astrophysical neutrinos in a maximum-likelihood analysis.
The combined event sample features high-statistics samples of shower-like and track-like events.
The data are fit in up to three observables: energy, zenith angle and event topology.
Assuming the astrophysical neutrino flux to be isotropic and to consist of equal flavors at Earth, the all-flavor spectrum with neutrino energies between 25~TeV and 2.8~PeV is well described by an unbroken power law with best-fit spectral index $-2.50\pm0.09$ and a flux at 100~TeV of $\left(6.7_{-1.2}^{+1.1}\right)\cdot10^{-18}\,\mathrm{GeV}^{-1}\mathrm{s}^{-1}\mathrm{sr}^{-1}\mathrm{cm}^{-2}$.
Under the same assumptions, an unbroken power law with index $-2$ is disfavored with a significance of 3.8~$\sigma$ ($p=0.0066\%$) with respect to the best fit.
This significance is reduced to 2.1~$\sigma$ ($p=1.7\%$) if instead we compare the best fit to a spectrum with index $-2$ that has an exponential cut-off at high energies.
Allowing the electron neutrino flux to deviate from the other two flavors, we find a $\nu_e$~fraction of $0.18\pm0.11$ at Earth.
The sole production of electron neutrinos, which would be characteristic of neutron-decay dominated sources, is rejected with a significance of 3.6~$\sigma$ ($p=0.014\%$).
\end{abstract}


\section{INTRODUCTION}
\label{sec:intro}
\setcounter{footnote}{0}
With the completion of the IceCube detector in December 2010, a decade-long journey towards building a kilometer-scale neutrino detector was finally concluded \citep[for a review, see e.g.][]{spiering2012}.
Using this instrument, we aimed to detect the first extraterrestrial neutrinos, and thus to explore territories previously inaccessible to astronomy.
These neutrinos have now been detected: In \citet{hese2yearpaper, hese3yearpaper}, the IceCube Collaboration presented the first evidence for a flux of high-energy extraterrestrial neutrinos, based on a search for events that start and deposit $\gtrsim30$~TeV inside the detection volume of IceCube.
Indications for this flux had already been found in earlier analyses performed on data taken during the construction phase of IceCube \citep{ic59numupaper, ic40cascpaper, schoenwald2013}.
More recently, the flux was measured down to deposited energies of 10~TeV \citep{ic7986hybridpaper}, and evidence for it was observed using $\nu_\mu$-induced tracks from the Northern hemisphere \citep{ic7986numupaper}.

Here, we present a maximum-likelihood analysis that is based on the event samples of all these analyses.
The samples include track-like events, induced by charged-current $\nu_\mu$ interactions, as well as shower-like events, induced by charged-current $\nu_e$ and $\nu_\tau$ interactions and all-flavor neutral-current interactions.\footnote{In $\approx$17\% of all charged-current $\nu_\tau$ interactions, a muon is produced in the tau decay, leading to a track-like instead of a shower-like signature. Furthermore, charged-current $\nu_\mu$ interactions may appear as shower-like if the muon leaves the detector unnoticed.}
In this analysis, we derive improved constraints on the spectrum and the flavor composition of the astrophysical neutrino flux.

In the remainder of this section, we summarize the phenomenology of astrophysical neutrinos and of background muons and neutrinos created in the atmosphere of the Earth, and we introduce the IceCube detector.
The searches for astrophysical neutrinos that are used for the combined analysis presented here are discussed in section~\ref{sec:datasets}.
The analysis method is explained in section~\ref{sec:method}, the results are given in section~\ref{sec:results}.
We conclude with a discussion of the results in section~\ref{sec:discussion}.

\subsection{Astrophysical Neutrinos}
Astrophysical neutrinos\footnote{Here and in the rest of this article, we imply also anti-neutrinos when we speak of neutrinos.} are created in interactions of high-energy cosmic rays with other massive particles or photons \citep{gaisser1995}.
The neutrinos, being electrically neutral and hence unaffected by cosmic magnetic fields, will travel in straight lines from their point of origin to Earth.
If the interactions happen close to the acceleration sites of the cosmic rays, they will thus reveal those.
Unlike gamma rays, which are created in the same processes, the neutrinos are also unlikely to be absorbed during their journey \citep[see e.g.][]{learned2000}.
Because of these properties, neutrinos are ideal messengers to study the sources of high-energy cosmic rays.
Astrophysical neutrinos carry information about these sources in their energy spectrum and flavor composition, even if their individual positions on the sky cannot be resolved yet \citep{hooper2003,choubey2009,lipari2007,laha2013}.
Candidate sources include active galactic nuclei \citep[e.g.][]{stecker1991,muecke2003}, gamma-ray bursts \citep[e.g.][]{waxman1997,guetta2004}, starburst galaxies \citep[e.g.][]{loeb2006}, and galaxy clusters \citep[e.g.][]{murase2008}, as well as galactic objects like supernova remnants or pulsar wind nebulae \citep[e.g.][]{bednarek2005,kistler2006,kappes2007}.

To first order, the energy spectrum of astrophysical neutrinos follows that of the cosmic rays at their acceleration sites.
If Fermi shock acceleration is the responsible mechanism, a power law spectrum $E^{-\gamma}$ with $\gamma\simeq2$ is expected \citep{gaisser1990}, although the details depend on the characteristics of the specific sources \citep[see e.g.][]{becker2008}.
Furthermore, the majority of the astrophysical neutrinos are expected to arise from the decay of pions created in cosmic-ray interactions, i.e. $\pi\rightarrow\mu+\nu_\mu$, followed by $\mu\rightarrow e+\nu_e+\nu_\mu$.
The flavor composition resulting from this decay chain is $\nu_e:\nu_\mu:\nu_\tau=1:2:0$.
Taking into account long-baseline neutrino oscillations, the flavor composition at Earth is different, approximately $1:1:1$ for this scenario \citep{learned1995, athar2006}.
This first order model of the energy spectrum and flavor composition has often been used as a benchmark scenario in the past.

Second order corrections to this benchmark model arise e.g. from muon energy losses \citep{kashti2005} and muon acceleration \citep{klein2013}, these effects can alter both the energy spectrum and flavor composition of the astrophysical neutrino flux.
While the energy spectrum is difficult to constrain by general arguments, there are two limiting scenarios for the flavor composition at the sources\footnote{Neglecting the production of tau neutrinos at the sources, which is a common assumption \citep{choubey2009}.}:
muon-damped sources, in which the high-energy neutrino flux is suppressed due to energy loss processes of the muons, and neutron-beam sources, in which the neutrinos are created from neutron rather than pion decays.
The flavor compositions at the source for these (idealized) scenarios are $0:1:0$ and $1:0:0$, respectively \citep{lipari2007}.
Using the neutrino oscillation parameters from \citet[][inverted hierarchy]{gonzalezgarcia2014}, the expected flavor transitions for the three source classes discussed here are $1:2:0 \rightarrow 0.93:1.05:1.02$ (pion-decay), $0:1:0 \rightarrow 0.19:0.43:0.38$ (muon-damped), and $1:0:0 \rightarrow 0.55:0.19:0.26$ (neutron-beam).

The implications of deviations from the benchmark scenario have been discussed by several authors.
Assuming extragalactic sources, \citet{winter2013} argues that a spectral index $\gamma\gtrsim2.3$ is more easily explained in photohadronic scenarios, i.e. by $p\gamma$-interactions as the origin of the neutrinos.
Similarly, \citet{murase2013} point out that if a hadronuclear origin ($pp$-interactions) in extragalactic sources is assumed, measurements of the diffuse extragalactic gamma-ray background \citep[see][]{fermidiffuse} imply $\gamma\lesssim2.1-2.2$ under certain conditions.
However, hadronuclear models that take into account the diffuse gamma-ray background as well as recent IceCube data \citep{ic7986hybridpaper}, which point towards a softer spectrum, have also been proposed, e.g.\ by \citet{senno2015}.
Other authors invoke spectral arguments to propose that the astrophysical neutrino flux is produced by sources within the Milky Way \citep{neronov2014,gaggero2015}.
Finally, the implications of flavor ratios different from the benchmark scenario have been discussed e.g. by \citet{beacom2003, lipari2007, vissani2013}.
\citet{bustamante2015} give an overview over the flavor compositions resulting from various standard and non-standard neutrino production and propagation scenarios.

\subsection{Atmospheric Backgrounds}
All relevant backgrounds to searches for high-energy astrophysical neutrinos are created in cosmic ray-induced air showers in the atmosphere of the Earth.
Of all the particles created in air showers, only muons and neutrinos can reach the IceCube detector.
Atmospheric muons constitute, by far, the most abundant background, triggering the detector at a rate of several~kHz.
Characteristically, they reach the detector from above the horizon and are first detected on the boundary of the instrumented volume.
Atmospheric neutrinos are created at a similar rate, but are detected much less frequently due to their small interaction probabilities.
They reach the detector from all directions, in particular also from below the horizon.
Atmospheric neutrinos that have passed through the Earth are free of muon background, but also hard to distinguish from astrophysical neutrinos.
In contrast, atmospheric neutrinos arriving from above the horizon are often accompanied by atmospheric muons \citep{gaisser2014}, which is never the case for astrophysical neutrinos.

At lower energies, the atmospheric neutrino flux is dominated by so-called \textit{conventional} atmospheric neutrinos from the decays of kaons and charged pions.
Since these particles are likely to interact with air molecules before they decay, the resulting neutrino flux differs from the original cosmic ray flux in its energy and zenith angle dependence: the energy spectrum is steeper (approximately $E^{-3.7}$) and the flux is enhanced towards the horizon \citep{gaisser1990}.
An additional contribution to the atmospheric neutrino flux is expected from the decays of heavy, short-lived hadrons containing a charm or bottom quark.
At the energies relevant here, these almost always decay before having the chance to interact, giving rise to a flux of \textit{prompt} atmospheric neutrinos.
This flux is predicted to follow that of the cosmic rays more closely, with an energy spectrum of approximately $E^{-2.7}$ and an isotropic zenith angle distribution.
It is, however, yet to be conclusively observed and predictions for its magnitude vary strongly \citep{bugaev1989, martin2003, enberg2008, bhattacharya2015}.

\subsection{The IceCube Detector}
The IceCube detector \citep{firstyearperformancepaper} consists of 5160 optical modules, installed in the ice underneath the geographic South Pole at depths between 1450 and 2450~meters.
The modules are deployed on 86 strings, where the vertical spacing between two modules on a string is approximately 17~meters and the horizontal spacing between two strings is approximately 125~meters.
Eight of the strings are part of a more densely instrumented region in the center of the detector, DeepCore, which serves as a low energy extension to IceCube \citep{deepcorepaper}.
Each optical module is a spherical glass housing that contains a $10''$-photomultiplier \citep{pmtpaper}, together with digitization electronics \citep{daqpaper}.
The IceCube Neutrino Observatory also includes a cosmic-ray detector called IceTop, which consists of 81 stations that are installed on the surface above the IceCube detector \citep{icetoppaper}.

IceCube detects neutrino interactions by measuring the Cherenkov radiation that is induced by the secondary particles created in the interaction.
The signature of a neutrino interaction in the detector depends on the flavor of the incoming neutrino as well as on the interaction type.
So-called \textit{tracks} arise from charged-current $\nu_\mu$ interactions, in which a muon is produced together with a hadronic particle shower at the interaction vertex.
At the relevant energies, the muon has a range of several kilometers, leading to an elongated, track-like signature.
Due to the long lever arm, the directional reconstruction of these events is very precise, with a typical median resolution of better than $1^\circ$ \citep{moonshadowpaper}.
Since the muon will typically leave the detector, and often also enter it from outside, only a lower limit can be determined for the neutrino energy.
On the other hand, charged-current $\nu_e$ and $\nu_\tau$ interactions as well as neutral current interactions of all flavors give rise to \textit{showers}.\footnote{At very high energies ($\gtrsim1$~PeV), the tau created in a charged-current $\nu_\tau$ interaction can travel sufficiently far ($\sim50$~m/PeV) before decaying such that two separated showers may be observed; one at the interaction vertex and one at the decay point of the tau. However, this ``double-bang'' signature has not been observed yet.}
The electromagnetic and hadronic particle showers produced in these interactions have dimensions that are typically much smaller than the detector spacing, so they appear as point sources of light.
As a result, the directional reconstruction of these events is worse than for tracks, typically $15^\circ$ \citep{energyrecopaper}.
For both event topologies, the energy deposited in the detector, which is a lower bound for the neutrino energy, can be reconstructed with a resolution of about 15\% \citep{energyrecopaper}.
For a track event, usually the differential energy loss or the total muon energy are estimated.
These quantities are correlated with the energy of the neutrino that produces the muon.
However, their absolute values should be interpreted with care \citep{energyrecopaper}.
Because the absolute values are unimportant here, the quantities are quoted in arbitrary units in this article.

The construction of the IceCube detector was finished in December 2010.
However, data had already been taken since 2005, after the installation of the first IceCube string.
Data used in this analysis were taken with configurations consisting of 40 strings (2008/2009), 59 strings (2009/2010), 79 strings (2010/2011) and the full 86-string configuration (since 2011).

\section{SEARCHES FOR HIGH-ENERGY NEUTRINOS WITH ICECUBE}
\label{sec:datasets}
Searches for high-energy astrophysical neutrinos have been performed ever since the beginning of IceCube data-taking, as before with data collected by IceCube's predecessor AMANDA \citep[see e.g.][]{amandanumupaper, ic22cascpaper}.
Here, we consider only more recent searches, and only those targeted at the TeV--PeV energy range.
We also restrict ourselves to searches for a diffuse flux of neutrinos, i.e. we do not include searches that solely aim to identify individual sources, since these typically contain a larger fraction of atmospheric muon background events.
This leaves a total of six searches, listed in Table~\ref{tab:analyses}.

The searches mainly differ in their strategy to suppress the large background of muons and neutrinos that are created in cosmic ray-induced air showers in the atmosphere of the Earth.
The important characteristics of such a strategy are the event topologies that are selected (listed in the table under ``Topology''), the use (or lack) of containment criteria (``Containment''), and the resulting neutrino energy and zenith angle ranges in which the search is sensitive to a flux of astrophysical neutrinos (``Energy range'', ``Zenith range'').
The table also lists the corresponding data taking period and the observables that are used in the maximum-likelihood analysis.
The individual searches and their background suppression strategies are briefly explained in the following sections, where we have classified them by topology into searches for track-like events (T1, T2), searches for shower-like events (S1, S2) and hybrid event searches, which select both kinds of events (H1, H2).
For more detailed information, we refer to the references listed in Table~\ref{tab:analyses}.
A table listing the fraction of neutrino interaction types that contribute to the individual samples can be found in Appendix \hyperlink{apx:a}{A}.

\begin{deluxetable*}{cccccccc}
  \tablecaption{Searches Combined in the Maximum-likelihood Analysis\label{tab:analyses}}
  \tablehead{\colhead{ID} & \colhead{Topology} & \colhead{Containment} & \colhead{\parbox{2.2cm}{\centering Energy range\tablenotemark{a}\\(TeV)}} & \colhead{\parbox{2.2cm}{\centering Zenith range\\(deg)}} & \colhead{\parbox{1.8cm}{\centering Data taking\\period}} & \colhead{Observables} & \colhead{Reference}}
  \startdata
    T1 & tracks & no & $>100$ & $90-180$ & $2009-2010$ & energy, zenith & 1\\
    T2 & tracks & no & $>100$ & $85-180$ & $2010-2012$ & energy, zenith & 2\\
    S1 & showers & yes & $>100$ & $0-180$ & $2008-2009$ & energy & 3\tablenotemark{b}\\
    S2 & showers & yes & $>30$ & $0-180$ & $2009-2010$ & energy & 4\tablenotemark{c}\\
    H1 & showers, tracks & yes & $>50$ & $0-180$ & $2010-2013$ & energy, zenith & 5,6\\
    H2 & showers, tracks & yes & $>20$ & $0-180$ & $2010-2012$ & energy, zenith, topology & 7
  \enddata
  \tablenotetext{a}{Refers to the approximate neutrino energy range in which the analysis is sensitive to astrophysical neutrinos, not the range of deposited energies selected.}
  \tablenotetext{b}{Samples ``Ia'' and ``Ib'' from the reference are used here.}
  \tablenotetext{c}{``Analysis A'' in the reference. Also see Appendix \hyperlink{apx:b}{B} for additional changes to the event selection made within this work.}
  \tablerefs{(1) \citet{ic59numupaper}; (2) \citet{ic7986numupaper}; (3) \citet{ic40cascpaper}; (4) \citet{schoenwald2013};\\
  (5) \citet{hese2yearpaper}; (6) \citet{hese3yearpaper}; (7) \citet{ic7986hybridpaper}}
\end{deluxetable*}

\subsection{Searches for Track-Like Events}
Searches of this kind select upward-going and horizontal $\nu_\mu$-induced tracks.
The good angular resolution of these events provides excellent discrimination against downward-going muon events, which can be reduced to a negligible level.
Because events with interaction vertices outside the instrumented volume can be selected, this strategy allows the selection of large neutrino samples of high purity.
It is however restricted by design to directions from which only neutrinos can reach the detector, i.e. the northern hemisphere and directions close to the horizon.
This implies that atmospheric neutrinos are an irreducible background to searches of this type and can only be distinguished from astrophysical neutrinos on a statistical basis.
Because atmospheric neutrinos are predominantly muon neutrinos at all relevant energies, a flux of astrophysical neutrinos will only be visible at very high energies $\gtrsim100$~TeV.
Neutrinos of these energies have a non-vanishing probability to be absorbed during their passage through the Earth \citep{gandhi1996}, searches for track-like events are therefore most sensitive in the direction of the horizon, where the amount of traversed matter is less compared to more vertical directions.

Here we use the neutrino event samples from two different event selections following this approach, performed on data taken in different periods.
The first one (``T1'') uses data taken in 2009/2010 with the 59-string configuration of IceCube while the second one (``T2'') uses two years of data, taken between 2010 and 2012 with the 79-string and the 86-string configuration of IceCube.
For both searches, energy and zenith angle information are available for the neutrino sample.
Note that we do not use the full event sample provided by the analyses, but restrict ourselves to the high-energy tail of the spectrum.
This degrades the ability to constrain the atmospheric neutrino spectrum (and thus, indirectly, the astrophysical spectrum) somewhat, but facilitates the consistent treatment of systematic uncertainties (which are difficult to account for in the high-statistics threshold region of the track-like searches) in the combined analysis presented here.

\subsection{Searches for Shower-Like Events}
These searches select shower-like events and are thus sensitive to all neutrino flavors.
Lacking the good angular resolution of tracks, searches of this type achieve background suppression by selecting events with a spherical topology, by requiring the interaction vertex to be contained within the instrumented volume of IceCube, and by vetoing events in which light is first detected on the outermost layer of optical modules.
This means that the sensitive analysis volume is smaller than the instrumented volume, but allows the selection of neutrinos coming from any direction.
However, reliable rejection of muon background events is only possible if there is a sufficiently high probability of detecting the incoming muon.
This typically leads to a contamination of residual muon background events at low energies, where muons may produce too little light to be rejected efficiently enough.

Event samples selected by two searches for shower-like events are included here, one performed on data taken in 2008/2009 with the 40-string configuration of IceCube (``S1'') and one performed on data taken in 2009/2010 with the 59-string configuration of IceCube (``S2'').
Note that the event sample of search S2 has been extended to lower energies compared to what was presented in \citet{schoenwald2013}, see Appendix \hyperlink{apx:b}{B} for further details.
Neither analysis originally accounted for the atmospheric self-veto effect, i.e. the possibility to reject atmospheric neutrinos due to atmospheric muons that accompany them \citep{schoenert2009}.
For the analysis presented here, the results were corrected for this effect based on a calculation by \citet{gaisser2014}; checks were made to ensure that systematic errors in the description of the effect for these event samples do not affect the results obtained here.
The search S1 provides two event samples with slightly different selection criteria, a low-energy sample (with deposited energies between 2~TeV and 100~TeV, ``S1a'') and a high-energy sample (with deposited energies $>100$~TeV, ``S1b'').
Both shower searches provide reconstructed energies for their respective event samples.

\subsection{Hybrid Event Searches}
The two hybrid searches considered here do not make any requirements on the topology of the selected events, which makes them sensitive to any kind of neutrino interaction and to all directions.
The first hybrid analysis (``H1'') utilizes the outermost layer of optical modules as a veto against incoming muons, similar to the two shower searches S1 and S2 described above.
To ensure a sufficiently high veto probability, it is additionally required that more than 6,000~photo-electrons, integrated over all optical modules, be recorded.
This value corresponds to a threshold of $\sim30$~TeV in deposited energy.
The second hybrid analysis (``H2'') employs a more complex veto algorithm that adapts to the energy of the event and additionally searches for early hits that could be caused by an incoming muon.
As a result, the threshold for this event selection is significantly lower, $\sim1$~TeV in deposited energy.
A residual muon background contamination is present in the event samples of both searches, H1 and H2.

\clearpage
Analysis H1 uses data taken in $2010-2013$ while analysis H2 uses data taken in $2010-2012$, both with the 79-string and 86-string configurations of IceCube.
For both event samples, energy and zenith angle information are available.
Analysis H2 additionally provides data on the topology of the selected events; those in which more than 10~photo-electrons can be associated with an outgoing muon track are classified as tracks, all others as showers.\footnote{This observable identifies $\sim$60\% ($\sim$80\%) of all charged-current $\nu_\mu$-interactions as tracks at 10~TeV (1~PeV).}
The atmospheric self-veto effect was implemented in the same way as for the shower searches (see above), based on \citet{gaisser2014}.

\subsection{Dataset Overlap}
\label{sec:overlap}
Data taken in one period is often subjected to multiple analyses, some of which may have overlapping event samples.
Specifically, among the analyses considered here, three (T2, H1, H2) use data taken between 2010 and 2012 and have event samples that overlap.
For the combined analysis presented here however, it is important that the different event samples be statistically independent, i.e. that individual events are not selected more than once.
To achieve this, we identify the overlap between the samples and remove the corresponding events from all but one of the samples, both in simulation and experimental data.
In particular, we remove events that start inside the instrumented volume of IceCube from the event sample of analysis T2 and we consider only events from the event sample of analysis H2 for which fewer than 6,000~photo-electrons were recorded.
With these adjustments, all event samples combined in the maximum-likelihood analysis are statistically independent.

\section{ANALYSIS METHOD}
\label{sec:method}
In order to obtain a global picture of the astrophysical neutrino flux measured with IceCube, we analyze data from different event selections in a single maximum-likelihood analysis.
From each of the searches for astrophysical neutrinos introduced in the previous section, we use the final selection level sample of experimental data events as well as corresponding samples of simulated neutrino events and, if applicable, muon events.
The data are binned in the available observables and analyzed together.
For each component contributing to the measured flux, probability density functions (PDFs) are obtained from the simulated event samples by weighting them to different models.
The sum of all PDFs is then compared to the distribution of experimental data.
In the maximum-likelihood procedure, the parameters of the models are varied until the best agreement with the data is achieved.
The models for the different background components and the astrophysical component are described in sections~\ref{sec:bg} and \ref{sec:signal}, followed by a description of the likelihood method (section~\ref{sec:llh}).

\subsection{Modeling of Background Components}
\label{sec:bg}
The relevant background components to all searches considered here are atmospheric neutrinos and atmospheric muons.
We use the same models to describe the neutrino background in all of the different searches.
The muon background is more specific and is hence modeled individually for each analysis.

\subsubsection{Atmospheric Neutrinos}
\label{sec:atmnu}
The models and parameters for the atmospheric neutrino components are summarized in Table~\ref{tab:bgmodels}.
For the conventional atmospheric neutrinos, we use the HKKMS model by \citet{honda2007} and for the prompt atmospheric neutrinos, we use the ERS model by \citet{enberg2008}.
Both models were slightly modified \citep[see][]{ic59numupaper} in order to be in accordance with measurements of the cosmic-ray spectrum in the energy range above 1~TeV; these modifications are based on the cosmic-ray composition model introduced in \citet{gaisser2012}.
The normalizations of both neutrino spectra ($\phi_{\rm{conv}}$, $\phi_{\rm{prompt}}$) are free parameters in the fit procedure.\footnote{An update of the ERS model was recently presented by \citet{bhattacharya2015}, who find a lower prompt flux estimate. The spectral shape of the modified ERS model used here agrees well with that of the new model though, so that, because the absolute normalization is a free fit parameter, it can still be used without being in contradiction with the new model.}

\begin{deluxetable}{ccc}
  \tablecaption{Neutrino Background Models.\label{tab:bgmodels}}
  \tablehead{\colhead{Component} & \colhead{Model} & \colhead{Parameters}}
  \startdata
    Atmospheric conventional $\nu$ & 1,3 & $\phi_{\mathrm{conv}}$\\
    Atmospheric prompt $\nu$ & 2,3 & $\phi_{\mathrm{prompt}}$
  \enddata
  \tablerefs{(1) \citet{honda2007};\\(2) \citet{enberg2008}; (3) \citet{gaisser2012}}
\end{deluxetable}

\subsubsection{Atmospheric Muons}
\label{sec:atmmu}
The contamination of atmospheric muon background in the event samples of the track-based analyses T1 and T2 is negligible \citep{ic59numupaper, ic7986numupaper}.
In order to estimate the residual muon background in the event samples of the other analyses, simulations of cosmic ray air showers with CORSIKA \citep{heck1998} were carried out.
For analysis S1, the expected muon contamination amounts to $\sim$57\% of the observed number of events in the low-energy sample (S1a) and $\sim$2\% in the high-energy sample (S1b) \citep{ic40cascpaper}; for analysis S2 the expected contamination is $\sim$12\% \citep{schoenwald2013}.
The contributions at high energies were determined by an extrapolation of the distribution at lower energies because of insufficient statistics at high energy.
The magnitude of the muon background in the event sample of analysis H1 was determined from experimental data and is expected to be $\sim$23\% of the observed number of events \citep{hese2yearpaper, hese3yearpaper}.
Finally, the expected contamination in the analysis H2 is $\sim$4\% \citep{ic7986hybridpaper}.
In all analyses, the estimation of the muon background is associated with large uncertainties.
Accordingly, the normalizations of the muon background contributions in each analysis are implemented as nuisance parameters, see section~\ref{sec:sys}.

\subsection{Modeling of the Astrophysical Component}
\label{sec:signal}
So far, the sources of the high-energy neutrino flux measured by IceCube have escaped identification \citep{hese3yearpaper, 4yearpspaper}.
Here, we therefore test different models for the astrophysical component, based on general assumptions about the neutrino flux.
These models are listed in Table~\ref{tab:models}.

The simplest model is the ``single power law'' model.
We assume that the astrophysical flux arrives isotropically from all directions, that its flavor-ratio at Earth is $\nu_e:\nu_\mu:\nu_\tau=1:1:1$, and that it can be described by a simple power law of the form

\begin{equation}\label{eq:flux}
  \Phi_\nu=\phi\cdot\left(\frac{E}{100\,\mathrm{TeV}}\right)^{-\gamma}\,.
\end{equation}
$\Phi_\nu$ denotes the all-flavor neutrino flux, $\phi$ its value at 100~TeV and $\gamma$ the power law spectral index.

The ``differential'' model is based on the same assumptions about isotropy and flavor composition, but models the astrophysical flux with nine independent basis functions, defined in nine logarithmically spaced energy intervals between 10~TeV and 10~PeV.
The normalizations $\phi_1-\phi_9$ of the basis functions are free fit parameters, while the energy spectrum in each interval is assumed to be $\propto E^{-2}$.
This model is similar to the procedure in \citet{hese3yearpaper, ic7986hybridpaper}.

In the ``north-south'' model, we relax the assumption of isotropy and allow for two independent astrophysical neutrino fluxes, one from the northern and one from the southern hemisphere (separated at declination $\delta=0^\circ$), both following a spectrum as defined in equation~(\ref{eq:flux}).
While this scenario might lack a good astrophysical motivation, it does allow us to describe the flux separately in two hemispheres that are affected by different detector-related systematic effects.
We refrain from testing more complex anisotropic models here because, having been selected for diffuse searches, the event samples used in this analysis currently do not contain unblinded right-ascension information.
We do however note that our simple north-south model could be sensitive to certain anisotropic scenarios, like e.g. an additional astrophysical component from the inner Galaxy.

Finally, the assumptions about the flavor composition are weakened in the ``2-flavor'' and ``3-flavor'' models.
In the 2-flavor model, the $\nu_\mu$ and $\nu_\tau$ flux are assumed to be equal at Earth, while the $\nu_e$ flux is allowed to deviate.
This relation is, in good approximation, true for any flavor composition at the source if standard neutrino oscillations transform the neutrino flux during propagation.\footnote{It is exactly true in the case of tribimaximal mixing, and valid up to $\sim$20\% for more realistic oscillation parameters.}
In the 3-flavor model, the normalizations of the fluxes of all three neutrino flavors are free parameters, allowing us to test for non-standard oscillation scenarios.
In both the 2-flavor and 3-flavor models, the fluxes of the individual neutrino flavors are assumed to have the same energy dependence, i.e. $\propto E^{-\gamma}$, where $\gamma$ is a free parameter.

\begin{deluxetable}{cc}
  \tablewidth{6cm}
  \tablecaption{Models for the Astrophysical Neutrino Flux.\label{tab:models}}
  \tablehead{\colhead{Model} & \colhead{Parameters}}
  \startdata
    single power law & $\phi$, $\gamma$\\[0.1cm]
    differential & $\phi_1-\phi_9$\\[0.1cm]
    north-south & $\phi_S$, $\gamma_S$, $\phi_N$, $\gamma_N$\\[0.1cm]
    2-flavor & $\phi_e$, $\phi_{\mu+\tau}$, $\gamma$\\[0.1cm]
    3-flavor & $\phi_e$, $\phi_\mu$, $\phi_\tau$, $\gamma$
  \enddata
\end{deluxetable}

\subsection{The Maximum-likelihood Method}
\label{sec:llh}
In the maximum-likelihood method, the agreement between experimental data and the simulated PDFs is established by means of the test statistic described in section~\ref{sec:ts}.
Section~\ref{sec:sys} lists the systematic uncertainties that we account for in the method. 
Finally, the calculation of \textit{p}-values for likelihood ratio and goodness-of-fit tests is explained in sections~\ref{sec:llhrtest} and \ref{sec:gof}, respectively.

\subsubsection{Test Statistic}
\label{sec:ts}
We use a binned Poisson-likelihood test statistic to compare the experimental data with the model predictions.
The general definition for the test statistic is

\begin{equation}\label{eq:ts}
  -2\ln L=-2\sum_i\ln\left(\frac{e^{-\nu_i}\nu_i^{n_i}}{n_i!}\right)\,,
\end{equation}
where $n_i$ and $\nu_i$ denote the number of observed and predicted events in bin $i$, respectively.

Systematic effects might change the PDFs that are fit to the data and could thus distort the results of the fit.
To avoid this, we have parametrized the impact on the PDFs of all relevant systematic effects and included them in the fit procedure as nuisance parameters.
For each nuisance parameter $\theta$ there is a prior, an additional, Gaussian-shaped penalty term in the likelihood function that penalizes deviations from the default central value $\theta^\ast$.
The width $\sigma[\theta]$ of the prior is based on the uncertainty associated with the systematic effect.
With $m$ nuisance parameters included, equation~(\ref{eq:ts}) now reads

\begin{equation}\label{eq:ts_mod}
  -2\ln L=-2\sum_i\ln\left(\frac{e^{-\nu_i}\nu_i^{n_i}}{n_i!}\right)+\,\sum_{j=1}^{m}\left(\frac{\theta_j-\theta_j^\ast}{\sigma[\theta_j]}\right)^2\,.
\end{equation}
The individual systematic effects considered as nuisance parameters are discussed in greater detail in the next section.

\subsubsection{Systematic Uncertainties}
\label{sec:sys}
The following systematic effects were included in the maximum-likelihood procedure (for a summary, see Table~\ref{tab:sys}):
\begin{itemize}
\item \textit{Cosmic-ray spectral index.}
Atmospheric neutrinos are produced by cosmic rays hitting the atmosphere, hence their energy spectrum depends on the cosmic-ray energy spectrum.
The uncertainty on the spectral index of this spectrum is implemented as a nuisance parameter that tilts the spectrum of atmospheric neutrinos by $\Delta\gamma_{\mathrm{cr}}$ relative to the default model.
Note that positive values of $\Delta\gamma_{\mathrm{cr}}$ correspond to \textit{softer} spectra.
We use an uncertainty on the cosmic-ray spectral index of 0.05 \citep[see e.g.][]{gaisser2012}.

\item \textit{Muon background normalization.}
The residual muon background was determined from simulations of cosmic-ray air showers and/or from data (see section~\ref{sec:atmmu}).
Due to the different methods used in the different individual analyses, systematic shifts can be uncorrelated between different analyses.
We have therefore implemented one nuisance parameter for each search that varies the normalization of the residual muon background in that search only.
Since the muon background is negligible for the track-based analyses T1 and T2, this amounts to a total of four muon background normalization parameters ($\phi_{\mu,\mathrm{S}1}$, $\phi_{\mu,\mathrm{S}2}$, $\phi_{\mu,\mathrm{H}1}$, $\phi_{\mu,\mathrm{H}2}$).
The width of the prior is 50\% for all of these parameters.

\item \textit{Energy scale shift.}
A shift of the energy scale could be introduced by several systematic effects.
Because the energy reconstruction is based on the number of detected photons, an uncertainty on the optical efficiency of the detector modules directly translates to an uncertainty on the energy scale.
An imperfect ice model, which describes the scattering and absorption of photons in the ice and is used both in simulation and event reconstruction, could also lead to a shift in the energy scale.
Since some analyses used older ice models than others, the effect could be different for the different analyses used here.
Therefore, one energy scale nuisance parameter $\phi_E$ was implemented for each analysis, shifting the energy scale for that analysis only.
We use a prior of 15\% on each each energy scale parameter.
\end{itemize}

We also checked for the impact of a deviation of the conventional atmospheric electron-to-muon neutrino ratio from the model prediction, which would result from a mis-modeling of the kaon-to-pion ratio in atmospheric air showers.
However, we found the impact on the parameters of the astrophysical neutrino flux to be negligible.

\begin{deluxetable}{cccc}
  \tablecaption{Systematic Uncertainties.\label{tab:sys}}
  \tablehead{\colhead{Uncertainty} & \colhead{Treatment} & \colhead{Parameter} & \colhead{Prior}}
  \startdata
    CR spectral index & global & $\Delta\gamma_{\mathrm{cr}}$ & $0\pm0.05$\\
    Atm. $\mu$ background & individual & $\phi_\mu$ & $1\pm0.5$\\
    Energy scale & individual & $\phi_E$ & $1\pm0.15$
  \enddata
  \tablecomments{The systematic uncertainties taken into account in this analysis.
  ``Treatment'' indicates whether this effect affects all analyses the same (``global'') or if the effect is treated different in each affected analysis (``individual'').\\
  }
\end{deluxetable}

\subsubsection{Likelihood Ratio Tests}
\label{sec:llhrtest}
Two different models H0 and H1 can be compared with \textit{likelihood ratio tests}.
For this, we use the quantity

\begin{equation}\label{eq:llhratio}
  -2\Delta\ln L=-2\ln L_{\mathrm{H0}}/L_{\mathrm{H1}}\,,
\end{equation}
which Wilk's theorem predicts to be $\chi^2$-distributed with $k$ degrees of freedom, where $k$ is the difference in the number of parameters between H1 and H0 \citep[chapter 38]{wilks1938, olive2014}.\footnote{This requires the models to be nested, i.e. each point in the parameter space of H0 can be accessed with the parameters of H1.}
The distribution can deviate from a $\chi^2$-distribution if the sample size is small or a parameter is close to a physical bound.
In this case, the exact distribution of $-2\Delta\ln L$ can be computed from toy Monte-Carlo experiments, generated from the best-fit parameter values of model H0.
The likelihood ratio test \textit{p}-value is then given by the percentage of toy experiments that have a larger value of $-2\Delta\ln L$ than observed in data.
The likelihood ratio test \textit{p}-values quoted in this article were obtained with this procedure.

We also employ likelihood ratio tests to determine confidence intervals on the model parameters.
For each parameter value, we perform a likelihood ratio test between the model with the parameter constrained to this value and the model with the parameter unconstrained.
This procedure is known as a profile likelihood scan.
In the $\chi^2$-approximation that we use here, the 68\% and 90\% confidence level intervals are given by the values at which $-2\Delta\ln L=1$ and $-2\Delta\ln L=2.71$, respectively.
Similarly, 2-dimensional confidence contours can be obtained by constraining two parameters simultaneously.

\subsubsection{Goodness-of-fit Tests}
\label{sec:gof}
With a slight modification, the quality of the fit of a single model can be assessed using the test statistic.
Similar to equation~(\ref{eq:llhratio}), we define

\begin{equation}
  -2\ln L/L_{\mathrm{sat}}=-2\sum_i\ln\left(e^{n_i-\nu_i}(\nu_i/n_i)^{n_i}\right)\,,
\end{equation}
where $L_{\mathrm{sat}}$ is the likelihood of the model that exactly predicts the observed outcome (i.e. $\nu_i\equiv n_i$).
After adding the same term as in equation~(\ref{eq:ts_mod}), this quantity is minimized by the same parameter values that minimize the test statistic defined there.
Moreover, in the large sample limit, the minimum value follows a $\chi^2$-distribution and as such allows the calculation of a goodness-of-fit \textit{p}-value \citep[chapter 38]{olive2014}.
Since our samples are not large, we determine the distribution of the modified test statistic by generating toy Monte-Carlo experiments from the best-fit parameter values.
Comparing this distribution with the value observed in experimental data gives the goodness-of-fit \textit{p}-value.

\section{RESULTS}
\label{sec:results}
In this section, we present the results of the maximum-likelihood analysis for the different models of the astrophysical neutrino flux given in section~\ref{sec:signal} and Table~\ref{tab:models}.

\subsection{Single Power Law Model}
The best-fit parameter values for the single power law model are listed in Table~\ref{tab:bestfit_singlepl}, including all nuisance parameters. Specifically, the all-flavor astrophysical neutrino flux is

\begin{equation}
  \phi=\left(6.7_{-1.2}^{+1.1}\right)\cdot10^{-18}\,\mathrm{GeV}^{-1}\mathrm{s}^{-1}\mathrm{sr}^{-1}\mathrm{cm}^{-2}
\end{equation}
at 100~TeV and the best-fit power law has a spectral index of

\begin{equation}
  \gamma=2.50\pm0.09\quad.
\end{equation}
This measurement is valid for neutrino energies between 25~TeV to 2.8~PeV.
This energy range was determined by successively removing events, ordered in energy, from the simulation data and repeating the fit with $\phi$ and $\gamma$ constrained to their best-fit values; its bounds denote the energies at which the test statistic worsens by $-2\Delta\ln L=1$.

We obtain a reasonable \textit{p}-value of 37.6\% from a goodness-of-fit test for this model.
A power law with a fixed index of $\gamma=2$ is disfavored with a significance of 3.8~$\sigma$ ($p=0.0066\%$) in a likelihood ratio test with respect to the model with a free spectral index.
We also tested a single power law model with a high-energy exponential cut-off as well as a model that consists of two isotropic astrophysical components, each described by a power law.
Neither model gave a better description of our data.
A power law with a fixed index of $\gamma=2$ and a high-energy exponential cut-off is still disfavored with a significance of 2.1~$\sigma$ ($p=1.7\%$) with respect to the model with a free spectral index.
The best-fit cut-off energy is $\left(1.6_{-0.7}^{+1.5}\right)\,\mathrm{PeV}$ for a fixed spectral index of $\gamma=2$, with a corresponding normalization of $\phi=(5.2_{-1.5}^{+1.9})\cdot10^{-18}\,\mathrm{GeV}^{-1}\mathrm{s}^{-1}\mathrm{sr}^{-1}\mathrm{cm}^{-2}$.
No cut-off is fitted for the best-fit spectral index of $\gamma=2.5$.

The likelihood analysis favors a prompt atmospheric component equal to zero, with a 90\% C.L. upper limit of 2.1 times the (modified) model prediction by \citet{enberg2008}.
This limit is slightly higher than that obtained in \citet{ic7986hybridpaper}; this is due partly to the way the data samples are combined here (cf. section~\ref{sec:overlap}) and to the different treatment of the energy scale uncertainty.

\begin{figure}
  \centering
  \includegraphics[width=.475\textwidth]{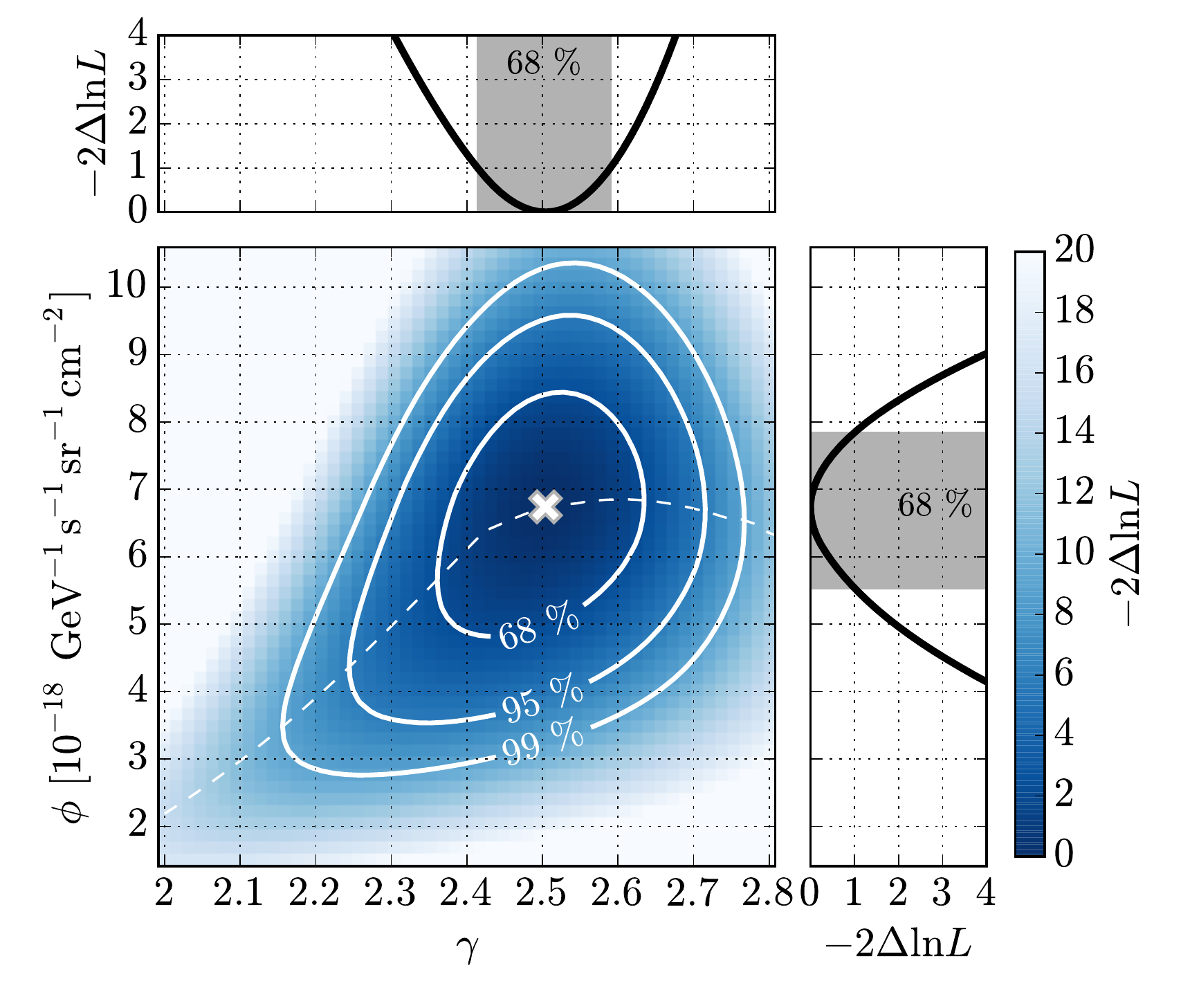}
  \caption{Profile likelihood scans around the best fit of the single power law model.
  The large panel displays a two-dimensional scan of the normalization $\phi$ and the spectral index $\gamma$ of the astrophysical neutrino flux; one-dimensional scans are shown in the small panels.
  The best-fit point is marked with ``$\times$'' in the large panel, the dashed line shows the conditional best-fit value of $\phi$ for each value of $\gamma$.\\
  }
  \label{fig:profllh_astro}
\end{figure}

\begin{figure}
  \centering
  \includegraphics[width=.475\textwidth]{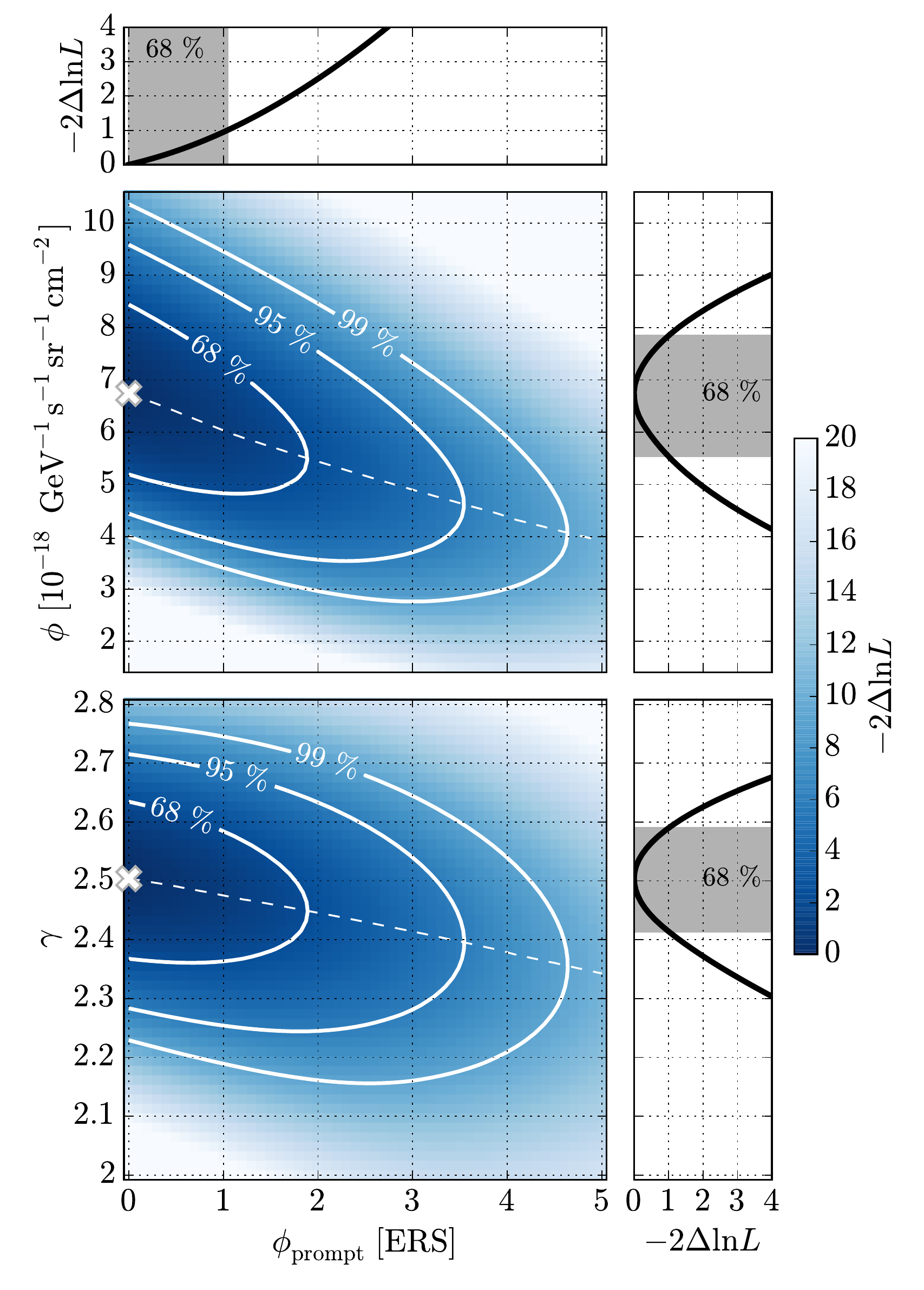}
  \caption{Profile likelihood scans around the best fit of the single power law model.
  The two large panels show two-dimensional scans of the normalization of the prompt atmospheric neutrino flux $\phi_\mathrm{prompt}$ and the normalization $\phi$ (upper panel) and spectral index $\gamma$ (lower panel) of the astrophysical neutrino flux, respectively; one-dimensional scans are shown in the small panels.
  See Figure \ref{fig:profllh_astro} for further description.
  }
  \label{fig:profllh}
\end{figure}

The correlation between the parameters $\phi$ and $\gamma$ is visualized in Figure~\ref{fig:profllh_astro}.
The behavior of these parameters as a function of the magnitude of the prompt component is shown in Figure~\ref{fig:profllh}.

\begin{deluxetable}{ccccc}
  \tablecaption{Best-Fit Parameter Values for the Single Power Law Model.\label{tab:bestfit_singlepl}}
  \tablehead{\colhead{Param.} & \colhead{Best fit} & \colhead{68\% C.L.} & \colhead{90\% C.L.} & \colhead{Pull}}
  \startdata
    $\phi_{\mathrm{conv}}$ & $1.10$ & $0.94-1.31$ & $0.87-1.49$ & $-$\\
    $\phi_{\mathrm{prompt}}$ & $0.00$ & $0.00-1.04$ & $0.00-2.11$ & $-$\\
    $\phi$ & $6.7$ & $5.5-7.8$ & $4.6-8.6$ & $-$\\
    $\gamma$ & $2.50$ & $2.41-2.59$ & $2.35-2.65$ & $-$\\[2pt]
    \tableline\\
    $\Delta\gamma_{\mathrm{cr}}$ & $0.017$ & $-0.008-0.041$ & $-0.023-0.057$ & $0.34$\\
    $\phi_{\mu,\mathrm{S}1}$ & $1.09$ & $0.72-1.51$ & $0.52-1.80$ & $0.18$\\
    $\phi_{\mu,\mathrm{S}2}$ & $0.84$ & $0.31-1.37$ & $0.00-1.71$ & $-0.32$\\
    $\phi_{\mu,\mathrm{H}1}$ & $1.12$ & $0.75-1.54$ & $0.56-1.84$ & $0.23$\\
    $\phi_{\mu,\mathrm{H}2}$ & $1.27$ & $0.94-1.61$ & $0.73-1.84$ & $0.54$\\
    $\phi_{E,\mathrm{S}1}$ & $0.95$ & $0.88-1.04$ & $0.84-1.12$ & $-0.34$\\
    $\phi_{E,\mathrm{S}2}$ & $1.00$ & $0.88-1.22$ & $0.83-1.32$ & $0.03$\\
    $\phi_{E,\mathrm{T}1}$ & $1.02$ & $0.95-1.09$ & $0.90-1.14$ & $0.10$\\
    $\phi_{E,\mathrm{T}2}$ & $1.05$ & $0.97-1.12$ & $0.93-1.17$ & $0.30$\\
    $\phi_{E,\mathrm{H}1}$ & $0.96$ & $0.88-1.06$ & $0.84-1.12$ & $-0.29$\\
    $\phi_{E,\mathrm{H}2}$ & $0.95$ & $0.86-1.04$ & $0.81-1.10$ & $-0.35$
  \enddata
  \tablecomments{$\phi$ is the value of the all-flavor neutrino flux at 100~TeV and is given in units of $10^{-18}\,\mathrm{GeV}^{-1}\mathrm{s}^{-1}\mathrm{sr}^{-1}\mathrm{cm}^{-2}$.
  $\phi_{\mathrm{conv}}$ and $\phi_{\mathrm{prompt}}$ are given as multiples of the model predictions (see Table \ref{tab:bgmodels}).
  ``Pull'' denotes the deviation of a nuisance parameter from its default value in units of the prior width $\sigma$.
  }
\end{deluxetable}

\begin{figure*}
  \centering
  \includegraphics[width=.8\textwidth]{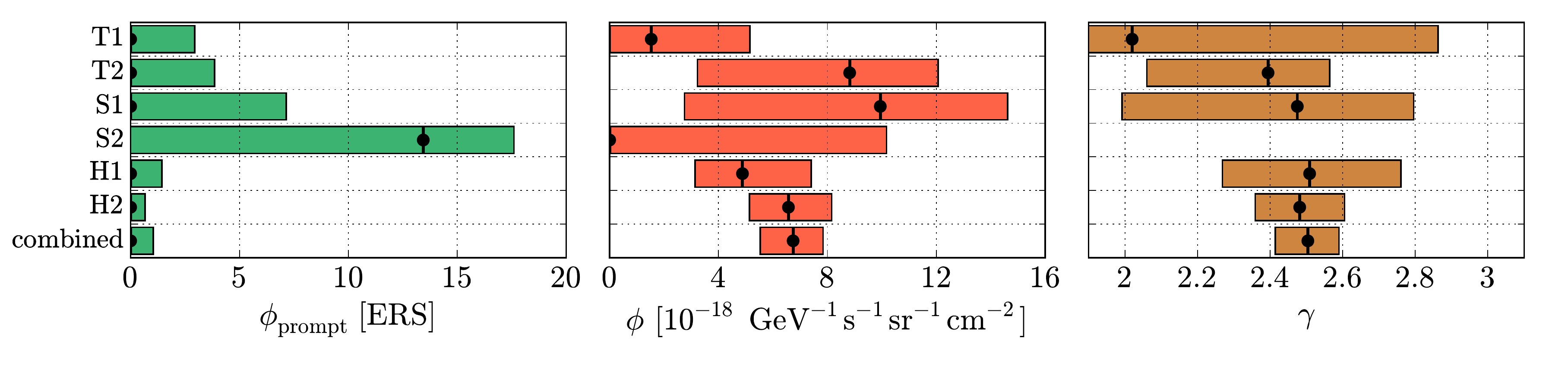}
  \caption{Results for the parameters $\phi_{\mathrm{prompt}}$ (left), $\phi$ (center), and $\gamma$ (right) in the single power law model.
  The last row (``combined'') shows the results of the combined analysis, while the other rows show the results obtained by applying the fit method from this work to the individual event samples.
  The best fit is marked by the black data point, the shaded area corresponds to the 68\% C.L. interval.
  }
  \label{fig:indiv_fits}
\end{figure*}

Figure~\ref{fig:indiv_fits} shows the results for the parameters $\phi_{\mathrm{prompt}}$, $\phi$, and $\gamma$ again (row ``combined''), together with results obtained from the application of the maximum-likelihood method described in this paper to the individual event samples.
For the individual fits, the event samples that were reduced to remove overlap (see section~\ref{sec:overlap}) were restored to their original size.
Furthermore, the normalization of the conventional atmospheric neutrino flux was treated as a nuisance parameter with a prior of 25\% around the model prediction for the fits on the event samples of analyses S1, S2, and H1, since this component is not well constrained by these samples alone.
A large prompt component instead of an astrophysical component is preferred in the fit of analysis S2 because these two components are close to degenerate if the only observable is the deposited energy and the astrophysical spectrum is steep.
Note that some of the results obtained with the individual samples differ slightly, although not significantly, from the originally published results (see references in Table~\ref{tab:analyses}).
For instance, a softer spectral index than presented in \citet{ic7986numupaper} is obtained for the sample T2 because only the high-energy data is used here; this difference is well within the uncertainty on this parameter.
The somewhat harder spectral index and lower normalization for sample H1 measured in \citet{hese3yearpaper} is a result of the reduced energy range ($>60$~TeV) used in the spectral fit there.

\begin{figure*}
  \centering
  \includegraphics[width=.999\textwidth]{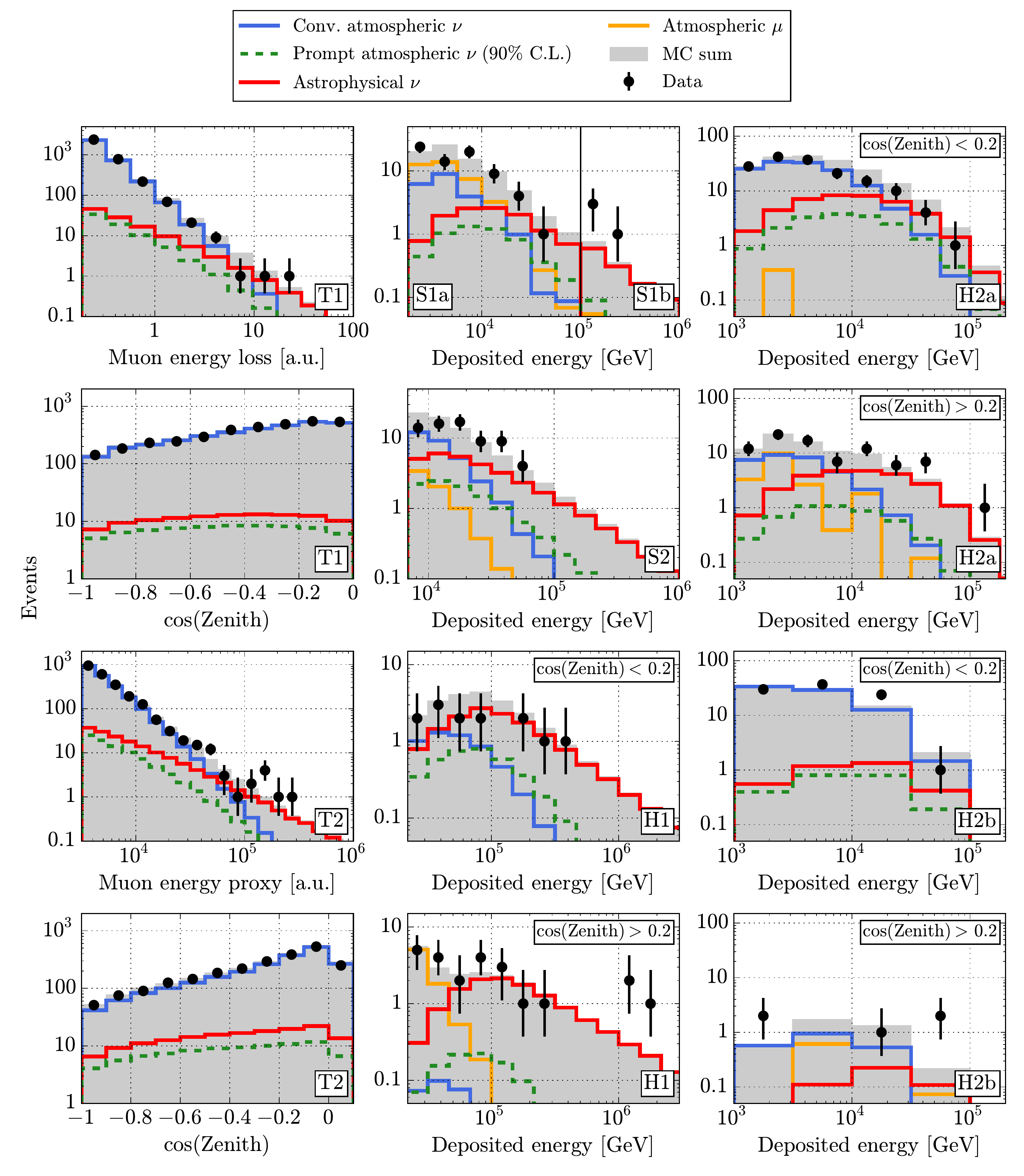}
  \caption{Experimental data and model predictions.
  The simulated data are weighted to the best-fit result of the single power law model.
  The event sample ID is indicated in the bottom right corner of each panel (cf. Table~\ref{tab:analyses}).
  For analysis S1, both event samples S1a and S1b are combined in one panel (top center).
  For analysis H2, the event sample is shown separately for showers (H2a) and tracks (H2b).
  If additional cuts were applied, they are indicated in the top right corner of the panel.
  The lines show the conventional atmospheric neutrino flux (blue), the astrophysical neutrino flux (red), and the atmospheric muon background (yellow).
  The 90\% C.L. upper limit on the prompt atmospheric neutrino flux is denoted by the dashed green line.
  The sum of all non-zero components is shown in gray.
  The black data points represent the experimental data, the errors bars denote 68\% C.L. intervals as defined in \citet{feldman1998}.
  }
  \label{fig:data_all}
\end{figure*}

Experimental and simulated data, weighted to the best-fit result, are shown in Figure~\ref{fig:data_all} for all event samples included in this analysis.
The data are projected onto observables that are used in the likelihood fit.
For the hybrid analyses H1 and H2, we show the energy distribution in two different zenith angle bins.

\begin{figure}
  \centering
  \includegraphics[width=.45\textwidth]{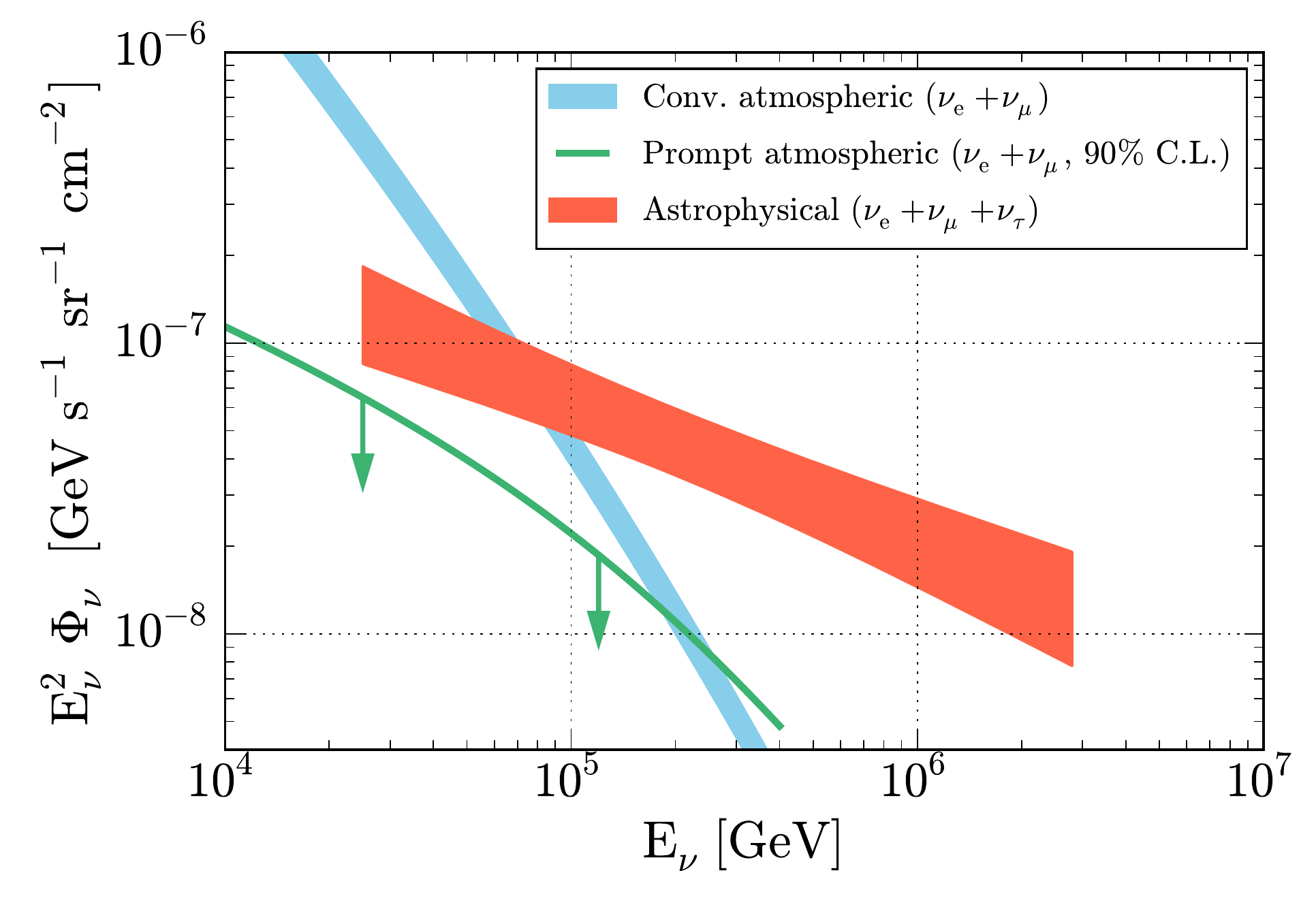}
  \caption{Best-fit neutrino spectra for the single power law model (all flavors combined).
  The blue and red shaded areas correspond to 68\% C.L. allowed regions for the conventional atmospheric and astrophysical neutrino flux, respectively.
  The prompt atmospheric flux is fitted to zero, we show the 90\% C.L. upper limit on this component instead (green line).
  }
  \label{fig:butterfly_atmos}
\end{figure}

Finally, the best-fit spectra for atmospheric and astrophysical neutrinos in the single power law model are shown in Figure~\ref{fig:butterfly_atmos}.

\subsection{Differential Model}
\begin{figure}
  \centering
  \includegraphics[width=.45\textwidth]{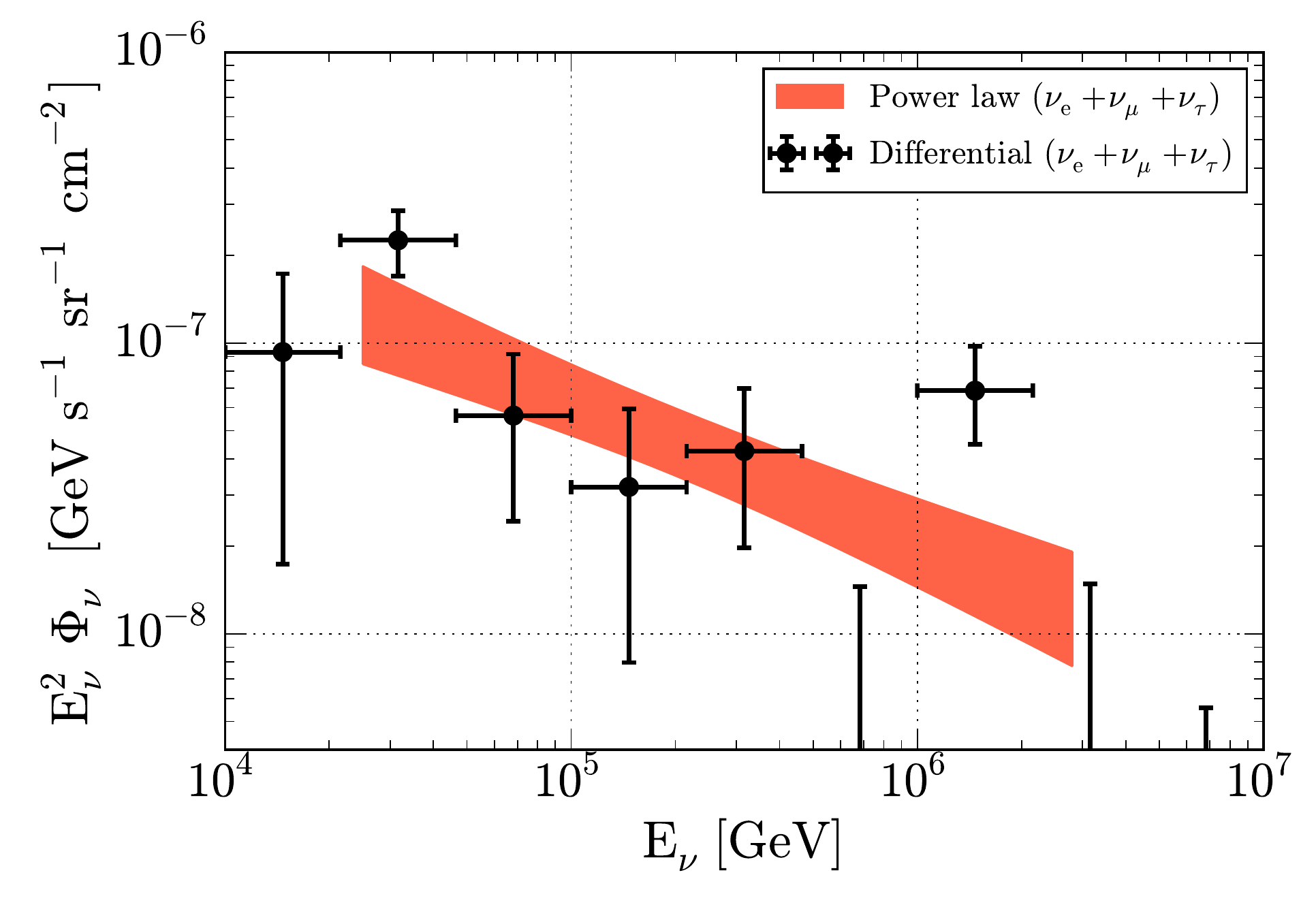}
  \caption{Best-fit astrophysical neutrino spectra (all flavors combined).
  The red shaded area corresponds to the 68\% C.L. allowed region for the single power law model (cf. Figure~\ref{fig:butterfly_atmos}).
  The black data points show the result of the differential model; the horizontal bars denote the bin width, the vertical error bars denote 68\% C.L. intervals.
  }
  \label{fig:butterfly_unfolding}
\end{figure}

Table~\ref{tab:bestfit_unfolding} lists the best-fit parameter values for the differential model.
$\phi_1$ through $\phi_9$ are the normalizations of the corresponding basis functions, each assumed to follow a spectrum $\propto E^{-2}$ and defined in logarithmically spaced energy intervals between 10~TeV and 10~PeV.
The resulting differential spectrum is shown in Figure~\ref{fig:butterfly_unfolding}, together with the single power law model (cf. previous section).

\subsection{North-South Model}
The best-fit values for the parameters of the astrophysical neutrino flux, separated into northern and southern hemisphere, are listed in Table~\ref{tab:bestfit_sep_ns}.
Most notable, the best-fit spectral index in the northern sky is $\gamma_N=2.0_{-0.4}^{+0.3}$, while in the southern sky it is $\gamma_S=2.56\pm0.12$.
This discrepancy with respect to the single power law model corresponds to a statistical \textit{p}-value of 13\% (1.1~$\sigma$).

\subsection{2-Flavor and 3-Flavor Model}
\begin{figure}
  \centering
  \includegraphics[width=.45\textwidth]{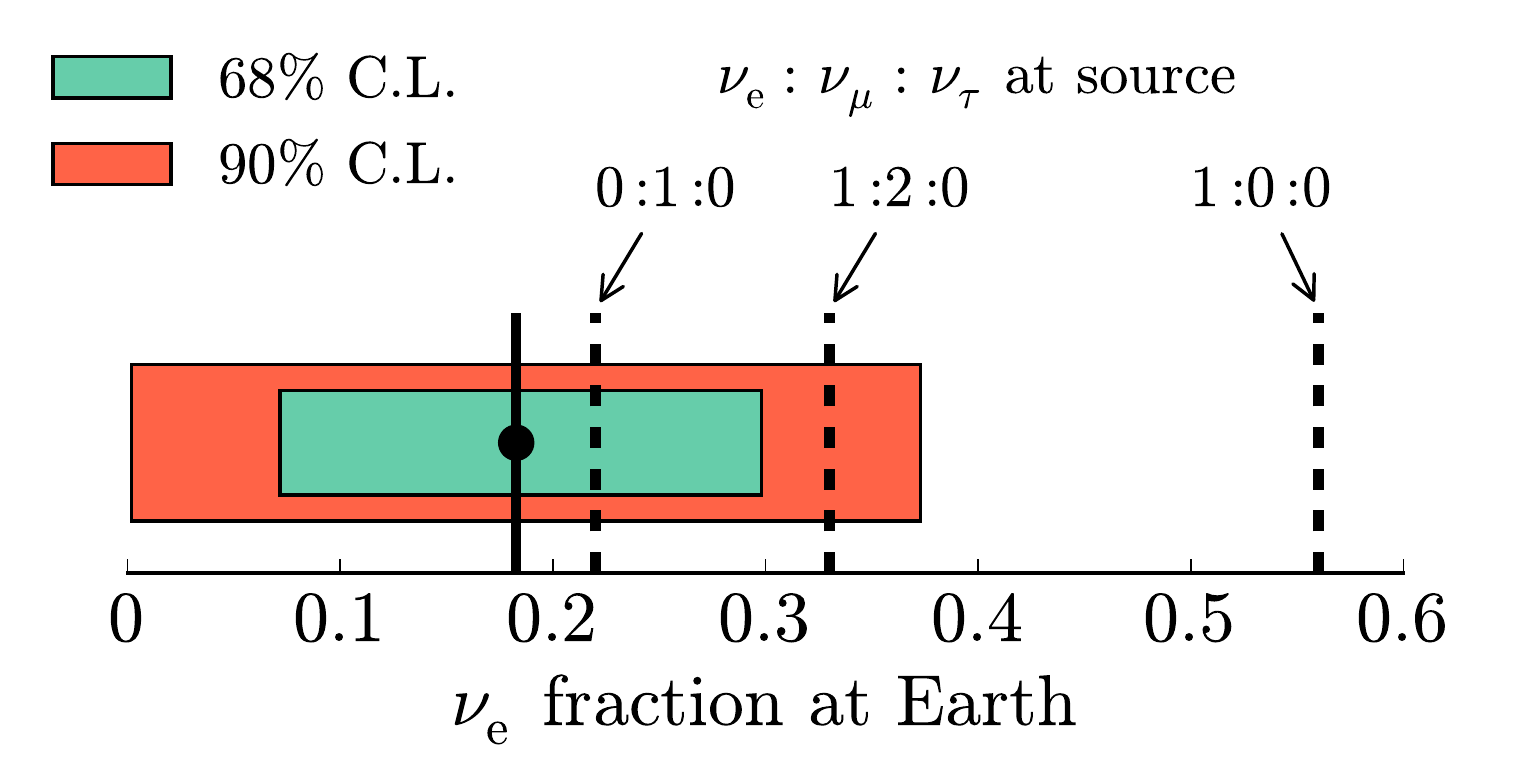}
  \caption{Electron neutrino fraction measured at Earth in the 2-flavor model.
  The black point denotes the best-fit value, the filled bands show the 68\% (green) and 90\% (red) C.L. intervals.
  The dashed lines mark electron neutrino fractions expected for different flavor compositions at the source, assuming tribimaximal neutrino mixing angles.
  }
  \label{fig:nue_fraction}
\end{figure}

For the 2-flavor model, the best-fit values for the astrophysical $\nu_e$ and $\nu_\mu+\nu_\tau$ flux are listed in Table~\ref{tab:bestfit_2flavor}.
The constraints on the astrophysical spectral index, which is assumed to be the same for both flavor components, are identical to those obtained in the single power law model.
Furthermore, the results on the flavor composition do not depend strongly on the value of this parameter.
We measure an electron neutrino fraction of the astrophysical neutrino flux of $0.18\pm0.11$ at Earth.
Figure~\ref{fig:nue_fraction} compares this value to fractions expected for different composition scenarios at the sources of the astrophysical flux.

\begin{deluxetable}{cccc}
  \tablecaption{Best-Fit Parameter Values for the Differential Model.\label{tab:bestfit_unfolding}}
  \tablehead{\colhead{Parameter} & \colhead{Best fit} & \colhead{68\% C.L.} & \colhead{90\% C.L.}}
  \startdata
    $\phi_1$ & $9.3$ & $1.7-17.3$ & $0.0-22.7$\\
    $\phi_2$ & $22.6$ & $17.0-28.5$ & $13.5-32.5$\\
    $\phi_3$ & $5.6$ & $2.4-9.2$ & $0.5-11.6$\\
    $\phi_4$ & $3.2$ & $0.8-5.9$ & $0.0-7.9$\\
    $\phi_5$ & $4.3$ & $2.0-7.0$ & $0.8-9.0$\\
    $\phi_6$ & $0.0$ & $0.0-1.5$ & $0.0-3.5$\\
    $\phi_7$ & $6.9$ & $4.5-9.7$ & $3.1-11.9$\\
    $\phi_8$ & $0.0$ & $0.0-1.5$ & $0.0-3.8$\\
    $\phi_9$ & $0.0$ & $0.0-0.6$ & $0.0-1.5$
  \enddata
  \tablecomments{$\phi_1-\phi_9$ are the all-flavor normalizations (in $E^2\Phi$) of the individual basis functions, defined in nine logarithmically spaced energy intervals between 10~TeV and 10~PeV.
  They are given in units of $10^{-8}\,\mathrm{GeV}\,\mathrm{s}^{-1}\mathrm{sr}^{-1}\mathrm{cm}^{-2}$.
  }
\end{deluxetable}

\begin{deluxetable}{cccc}
  \tablecaption{Best-Fit Parameter Values for the North-South Model.\label{tab:bestfit_sep_ns}}
  \tablehead{\colhead{Parameter} & \colhead{Best fit} & \colhead{68\% C.L.} & \colhead{90\% C.L.}}
  \startdata
    $\phi_N$ & $2.1$ & $0.5-5.0$ & $0.1-7.3$\\
    $\gamma_N$ & $2.0$ & $1.6-2.3$ & $1.2-2.5$\\
    $\phi_S$ & $6.8$ & $5.3-8.4$ & $4.4-9.5$\\
    $\gamma_S$ & $2.56$ & $2.44-2.67$ & $2.36-2.75$
  \enddata
  \tablecomments{$\phi_N$ and $\phi_S$ are the all-flavor neutrino fluxes at 100~TeV in the northern and southern sky, respectively; $\gamma_N$ and $\gamma_S$ are the corresponding spectral indices.
  The fluxes are given in units of $10^{-18}\,\mathrm{GeV}^{-1}\mathrm{s}^{-1}\mathrm{sr}^{-1}\mathrm{cm}^{-2}$.
  }
\end{deluxetable}

\begin{deluxetable}{cccc}
  \tablecaption{Best-Fit Parameter Values for the 2-Flavor Model.\label{tab:bestfit_2flavor}}
  \tablehead{\colhead{Parameter} & \colhead{Best fit} & \colhead{68\% C.L.} & \colhead{90\% C.L.}}
  \startdata
    $\phi_e$ & $1.3$ & $0.5-2.1$ & $0.0-2.6$\\
    $\phi_{\mu+\tau}$ & $5.6$ & $4.4-6.9$ & $3.7-7.8$
  \enddata
  \tablecomments{$\phi_e$ and $\phi_{\mu+\tau}$ are the $\nu_e$ and $\nu_\mu+\nu_\tau$ flux at 100~TeV, respectively.
  Both are given in units of $10^{-18}\,\mathrm{GeV}^{-1}\mathrm{s}^{-1}\mathrm{sr}^{-1}\mathrm{cm}^{-2}$.\\
  }
\end{deluxetable}

\begin{figure}
  \centering
  \includegraphics[width=.45\textwidth]{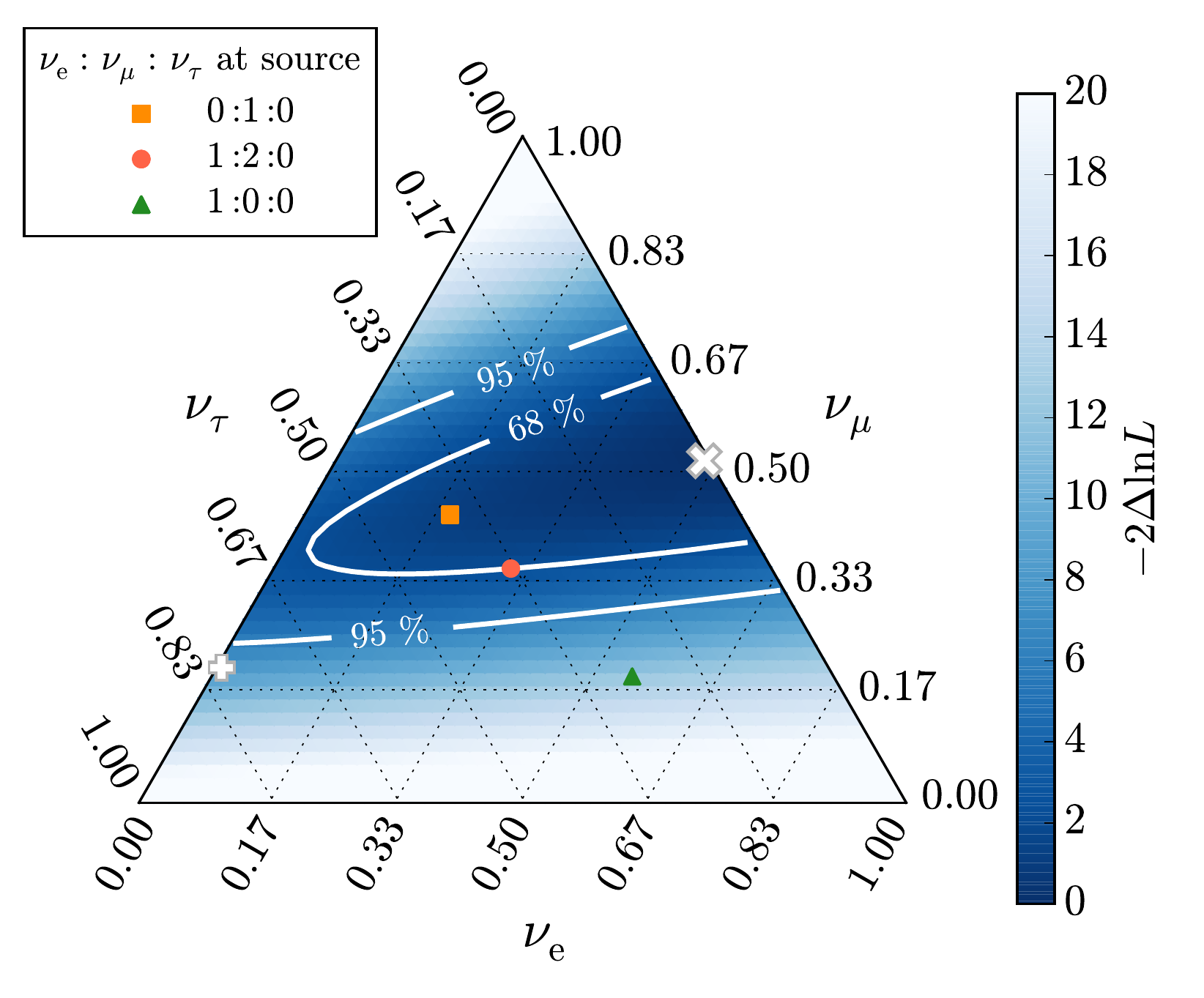}
  \caption{Profile likelihood scan of the flavor composition at Earth.
  Each point in the triangle corresponds to a ratio $\nu_e:\nu_\mu:\nu_\tau$ as measured on Earth, the individual contributions are read off the three sides of the triangle.
  The best-fit composition is marked with ``$\times$'', 68\% and 95\% confidence regions are indicated.
  The ratios corresponding to three flavor composition scenarios at the sources of the neutrinos, computed using the oscillation parameters in \citet[][inverted hierarchy]{gonzalezgarcia2014}, are marked by the square ($0:1:0$), circle ($1:2:0$), and triangle ($1:0:0$), respectively.
  The best-fit composition obtained in an earlier IceCube analysis of the flavor composition \citep{flavorpaper} is marked with a ``+''.\\
  }
  \label{fig:flavor_triangle}
\end{figure}

Finally, Table~\ref{tab:bestfit_3flavor} gives the best-fit values for the astrophysical fluxes in the 3-flavor model.
As in the 2-flavor model, we find that the results do not change significantly when varying the spectral index of the astrophysical flux within its uncertainties.
Note that none of the analyses considered here positively identifies charged-current tau neutrino interactions, which, depending on the tau decay mode, have a signature very similar to that of charged-current electron neutrino ($\sim$83\%) or muon neutrino ($\sim$17\%) interactions.
For this reason, the $\nu_\tau$ flux is easily conflated with the $\nu_e$ and $\nu_\mu$ components in the 3-flavor model.
Figure~\ref{fig:flavor_triangle} shows a profile likelihood scan of the flavor composition as measured on Earth, again compared to ratios expected for different source composition scenarios.
Performing likelihood ratio tests between these points and the best-fit hypothesis, we find \textit{p}-values of 55\% ($0:1:0$), 27\% ($1:2:0$), and 0.014\% ($1:0:0$), respectively.
Hence, our data are compatible with a pure muon neutrino composition and the generic pion-decay composition at the source, but reject a composition consisting purely of electron neutrinos at the source with a significance of 3.6~$\sigma$.

\begin{deluxetable}{cccc}
  \tablecaption{Best-Fit Parameter Values for the 3-Flavor Model.\label{tab:bestfit_3flavor}}
  \tablehead{\colhead{Parameter} & \colhead{Best fit} & \colhead{68\% C.L.} & \colhead{90\% C.L.}}
  \startdata
    $\phi_e$ & $2.9$ & $1.4-3.6$ & $0.0-4.2$\\
    $\phi_\mu$ & $3.0$ & $2.4-3.7$ & $2.1-4.2$\\
    $\phi_\tau$ & $0.0$ & $0.0-2.3$ & $0.0-5.0$
  \enddata
  \tablecomments{$\phi_e$, $\phi_\mu$, and $\phi_\tau$ are the $\nu_e$, $\nu_\mu$, and $\nu_\tau$ flux at 100~TeV, respectively.
  All are given in units of $10^{-18}\,\mathrm{GeV}^{-1}\mathrm{s}^{-1}\mathrm{sr}^{-1}\mathrm{cm}^{-2}$.
  }
\end{deluxetable}

\section{DISCUSSION}
\label{sec:discussion}
We have presented results of a maximum-likelihood analysis that is based on the combination of event samples that were selected by six different studies designed to measure an astrophysical neutrino flux with IceCube.
Assuming that the astrophysical flux is isotropic and has a flavor ratio of $\nu_e:\nu_\mu:\nu_\tau=1:1:1$ at Earth, we found that the combined data are well described by a power law with spectral index $-2.50\pm0.09$ between neutrino energies of 25~TeV and 2.8~PeV.
These are the most precise constraints on the spectrum of astrophysical neutrinos obtained so far.
With present statistics, we found no evidence that more complex spectral shapes or features, such as a high-energy cut-off or multiple spectral components, are required to describe the data.
However, astrophysically relevant deviations from a single power law are also not ruled out by the current data.
Under the same assumptions of isotropy and flavor equality, we exclude a spectrum $\propto E^{-2}$, which is the benchmark prediction of the Fermi shock acceleration model \citep[e.g.][]{gaisser1990}, with a significance of 3.8~$\sigma$ ($p=0.0066\%$).
Correspondingly, a spectrum $\propto E^{-2}$ with an additional high-energy exponential cut-off is disfavored with a significance of 2.1~$\sigma$ ($p=1.7\%$).

\citet{murase2013} pointed out that a soft spectrum ($\gamma\gtrsim2.1-2.2$) of the astrophysical neutrino flux might be difficult to accommodate with the observed extragalactic gamma-ray background \citep{fermidiffuse}, if the entire neutrino flux is produced in pp-interactions in extragalactic sources that are transparent to gamma rays and distributed according to the star-formation rate.
The soft spectrum obtained in this analysis could therefore be a first indication that at least part of the flux originates from other source classes, or that its spectral shape is more complex than assumed here.

While the analysis favors the absence of a prompt atmospheric neutrino component, a contribution to the flux at the level of the prediction by \citet{enberg2008} is still allowed and would not alter the spectrum of the astrophysical component significantly, as demonstrated in Figure~\ref{fig:profllh}.
Figure~\ref{fig:indiv_fits} shows that the results are compatible with those found in the individual studies; differences to the originally published results are mainly due to different energy ranges used in the analysis.

The strength of the astrophysical signal in different energy intervals is shown in the differential spectrum in Figure~\ref{fig:butterfly_unfolding}.
This spectrum suggests that it is mostly events with energies around 30~TeV that are responsible for the soft spectrum obtained in the analysis here.
In fact, a previous analysis \citep{hese3yearpaper} that did not include data at these energies yielded a harder spectral index of $-2.3\pm0.3$, but with larger uncertainties.
The result is compatible with the one obtained here.\footnote{We have established the compatibility in a separate fit without the corresponding data set, i.e. without sample H1. The 68\% uncertainty interval for the spectral index obtained in this fit ($-2.45\pm0.10$) overlaps with that obtained in \citet{hese3yearpaper}.}

We have tested the hypothesis of isotropy by fitting a model with two astrophysical components, one in the northern and one in the southern sky.
Compared to the all-sky result, the fit prefers a harder spectrum ($E^{-(2.0_{-0.4}^{+0.3})}$) in the northern sky and a slightly softer spectrum ($E^{-2.56\pm0.12}$) in the southern sky with a significance of 1.1~$\sigma$ ($p=13\%$).
The result is not conclusive; the discrepancy could be caused by a statistical fluctuation or by an additional component that is present in only one of the hemispheres (either an unmodeled background component, or e.g. a component from the inner Galaxy, although a single point source of the required strength to create the anisotropy anywhere in that region has already been excluded \citep{antarespspaper}).
Further analysis including right-ascension information will be helpful in testing the hypothesis of isotropy in the future.

Finally, we performed a measurement of the flavor composition of the astrophysical neutrino flux.
In a first test, we have measured the electron neutrino fraction at Earth in a tribimaximal mixing scenario, with equal $\nu_\mu$ and $\nu_\tau$ fluxes at Earth.
The best-fit fraction is $0.18\pm0.11$, a value compatible with the fractions expected from pion-decay sources (0.33) and muon-damped sources (0.22), but incompatible with that expected from neutron-beam sources (0.56), see Figure~\ref{fig:nue_fraction}.
In a second, more general test, we allow the normalizations of all three flavor components to vary independently and compare the result to compositions expected for different astrophysical scenarios in Figure~\ref{fig:flavor_triangle}.
In agreement with the first test, we find that pion-decay sources and muon-damped sources are well compatible with our data, while neutron-beam sources are disfavored with a significance of 3.6~$\sigma$ ($p=0.014\%$).
We do not find indications for non-standard oscillation scenarios.

Previous measurements of the flavor composition were presented by \citet{mena2014} and \citet{palomaresruiz2015} \citep[based on event sample H1, presented in][]{hese3yearpaper}, and by \citet{palladino2015}, \citet{pagliaroli2015}, and \citet{flavorpaper} (based on event samples that were extended with respect to H1, respectively).
With respect to these measurements, the constraints presented here are significantly improved; we attribute this to the fact that the combined event sample analyzed here contains a significant number of shower events as well as track events.
Though the best-fit flavor composition obtained in \citet{flavorpaper} (white ``+'' in Figure~\ref{fig:flavor_triangle}) lies outside the 95\% C.L. region, the 68\% C.L. region obtained here is completely contained within that obtained in the previous work, demonstrating the compatibility of the two results.
Because neither analysis was designed to identify tau neutrinos, a degeneracy with respect to the $\nu_\tau$-fraction is observed in both, the slight preference towards a smaller $\nu_\tau$-contribution found here is likely connected to the slight differences in the energy distributions of the three neutrino flavors.
In future, the identification of tau neutrinos will enable us to place stronger constraints on the flavor composition of the astrophysical neutrino flux.

\acknowledgments
We acknowledge the support from the following agencies:
U.S. National Science Foundation-Office of Polar Programs,
U.S. National Science Foundation-Physics Division,
University of Wisconsin Alumni Research Foundation,
the Grid Laboratory Of Wisconsin (GLOW) grid infrastructure at the University of Wisconsin - Madison, the Open Science Grid (OSG) grid infrastructure;
U.S. Department of Energy, and National Energy Research Scientific Computing Center,
the Louisiana Optical Network Initiative (LONI) grid computing resources;
Natural Sciences and Engineering Research Council of Canada,
WestGrid and Compute/Calcul Canada;
Swedish Research Council,
Swedish Polar Research Secretariat,
Swedish National Infrastructure for Computing (SNIC),
and Knut and Alice Wallenberg Foundation, Sweden;
German Ministry for Education and Research (BMBF),
Deutsche Forschungsgemeinschaft (DFG),
Helmholtz Alliance for Astroparticle Physics (HAP),
Research Department of Plasmas with Complex Interactions (Bochum), Germany;
Fund for Scientific Research (FNRS-FWO),
FWO Odysseus programme,
Flanders Institute to encourage scientific and technological research in industry (IWT),
Belgian Federal Science Policy Office (Belspo);
University of Oxford, United Kingdom;
Marsden Fund, New Zealand;
Australian Research Council;
Japan Society for Promotion of Science (JSPS);
the Swiss National Science Foundation (SNSF), Switzerland;
National Research Foundation of Korea (NRF);
Danish National Research Foundation, Denmark (DNRF)

\hypertarget{apx:a}{\mbox{}}
\section*{APPENDIX A\\TABLE OF INTERACTION TYPES}
Table~\ref{tab:inttypes} lists the fraction of neutrino interaction types that contribute to the event samples introduced in section~\ref{sec:datasets}.

\begin{deluxetable*}{ccccccccccccccccc}
\tablecaption{Fraction of Interaction Types per Event Sample\label{tab:inttypes}}
\tablehead{\colhead{} & \colhead{} & \multicolumn{7}{c}{$10\,\mathrm{TeV}<E_\nu<100\,\mathrm{TeV}$} & \colhead{} & \multicolumn{7}{c}{$100\,\mathrm{TeV}<E_\nu<1\,\mathrm{PeV}$}\\
           \colhead{} & \colhead{} & \multicolumn{2}{c}{$\nu_e$} & \multicolumn{2}{c}{$\nu_\mu$} & \multicolumn{2}{c}{$\nu_\tau$} & \colhead{} & \colhead{} & \multicolumn{2}{c}{$\nu_e$} & \multicolumn{2}{c}{$\nu_\mu$} & \multicolumn{2}{c}{$\nu_\tau$} & \colhead{}\\
           \colhead{ID} & \colhead{} & \colhead{CC} & \colhead{NC} & \colhead{CC} & \colhead{NC} & \colhead{CC} & \colhead{NC} & \colhead{GR} & \colhead{} & \colhead{CC} & \colhead{NC} & \colhead{CC} & \colhead{NC} & \colhead{CC} & \colhead{NC} & \colhead{GR}}
\startdata
\cutinhead{Conventional Neutrinos}
  T1 & & 0.0 & 0.0 & 100.0 & 0.0 & 0.0 & 0.0 & 0.0 & & 0.0 & 0.0 & 100.0 & 0.0 & 0.0 & 0.0 & 0.0\\
  T2\tablenotemark{a} & & 0.0 & 0.0 & 100.0 & 0.0 & 0.0 & 0.0 & 0.0 & & 0.0 & 0.0 & 100.0 & 0.0 & 0.0 & 0.0 & 0.0\\
  S1 & & 7.7 & 1.5 & 48.8 & 42.0 & 0.0 & 0.0 & 0.0 & & 5.2 & 1.5 & 40.8 & 52.5 & 0.0 & 0.0 & 0.0\\
  S2 & & 17.0 & 1.6 & 32.7 & 48.7 & 0.0 & 0.0 & 0.0 & & 8.8 & 2.3 & 30.2 & 58.7 & 0.0 & 0.0 & 0.0\\
  H1 & & 15.4 & 0.6 & 65.7 & 18.3 & 0.0 & 0.0 & 0.0 & & 6.9 & 0.7 & 73.1 & 19.3 & 0.0 & 0.0 & 0.0\\
  H2a\tablenotemark{a} & & 13.6 & 1.9 & 30.5 & 54.0 & 0.0 & 0.0 & 0.0 & & 2.4 & 2.5 & 23.2 & 71.9 & 0.0 & 0.0 & 0.0\\
  H2b\tablenotemark{a} & & 0.0 & 0.0 & 100.0 & 0.0 & 0.0 & 0.0 & 0.0 & & 0.0 & 0.0 & 99.2 & 0.8 & 0.0 & 0.0 & 0.0\\
\cutinhead{Astrophysical Neutrinos ($E^{-2.5}$)}
  T1 & & 0.0 & 0.0 & 96.5 & 0.0 & 3.5 & 0.0 & 0.0 & & 0.0 & 0.0 & 91.7 & 0.0 & 8.3 & 0.0 & 0.0\\
  T2\tablenotemark{a} & & 0.0 & 0.0 & 97.4 & 0.0 & 2.6 & 0.0 & 0.0 & & 0.0 & 0.0 & 96.9 & 0.0 & 3.1 & 0.0 & 0.0\\
  S1 & & 43.1 & 9.0 & 8.0 & 7.0 & 24.7 & 8.1 & 0.1 & & 34.2 & 10.3 & 7.8 & 11.0 & 26.7 & 9.7 & 0.3\\
  S2 & & 46.3 & 5.9 & 3.9 & 5.9 & 31.9 & 6.0 & 0.1 & & 35.9 & 9.7 & 4.7 & 9.4 & 29.6 & 10.4 & 0.3\\
  H1 & & 58.1 & 2.8 & 9.9 & 2.7 & 23.6 & 2.7 & 0.2 & & 37.8 & 4.9 & 19.4 & 5.0 & 27.4 & 5.1 & 0.4\\
  H2a\tablenotemark{a} & & 41.0 & 6.6 & 4.5 & 6.5 & 33.7 & 7.5 & 0.2 & & 12.2 & 15.2 & 5.0 & 15.1 & 34.0 & 18.4 & 0.1\\
  H2b\tablenotemark{a} & & 0.1 & 0.0 & 90.7 & 0.0 & 9.1 & 0.1 & 0.0 & & 0.3 & 0.1 & 79.6 & 0.8 & 18.5 & 0.7 & 0.0
\enddata
\tablecomments{Fraction, in \%, of neutrino interaction types contributing to the individual event samples.
CC = charged-current interaction; NC = neutral-current interaction; GR = Glashow resonance ($\bar{\nu}_e+e^-\rightarrow W^-$).
The top table shows numbers for a spectrum weighted to the conventional atmospheric neutrino flux of \cite{honda2007}; the bottom table is for an astrophysical power-law spectrum with index $-2.5$ and flavor composition $\nu_e:\nu_\mu:\nu_\tau=1:1:1$.
}
\tablenotetext{a}{Note that these event samples were adjusted to remove overlap between samples T2, H1, and H2, see section~\ref{sec:overlap}.\\}
\end{deluxetable*}

\hypertarget{apx:b}{\mbox{}}
\section*{APPENDIX B\\CHANGES MADE TO THE EVENT SELECTION OF SAMPLE S2}
\begin{figure}
  \centering
  \includegraphics[width=.45\textwidth]{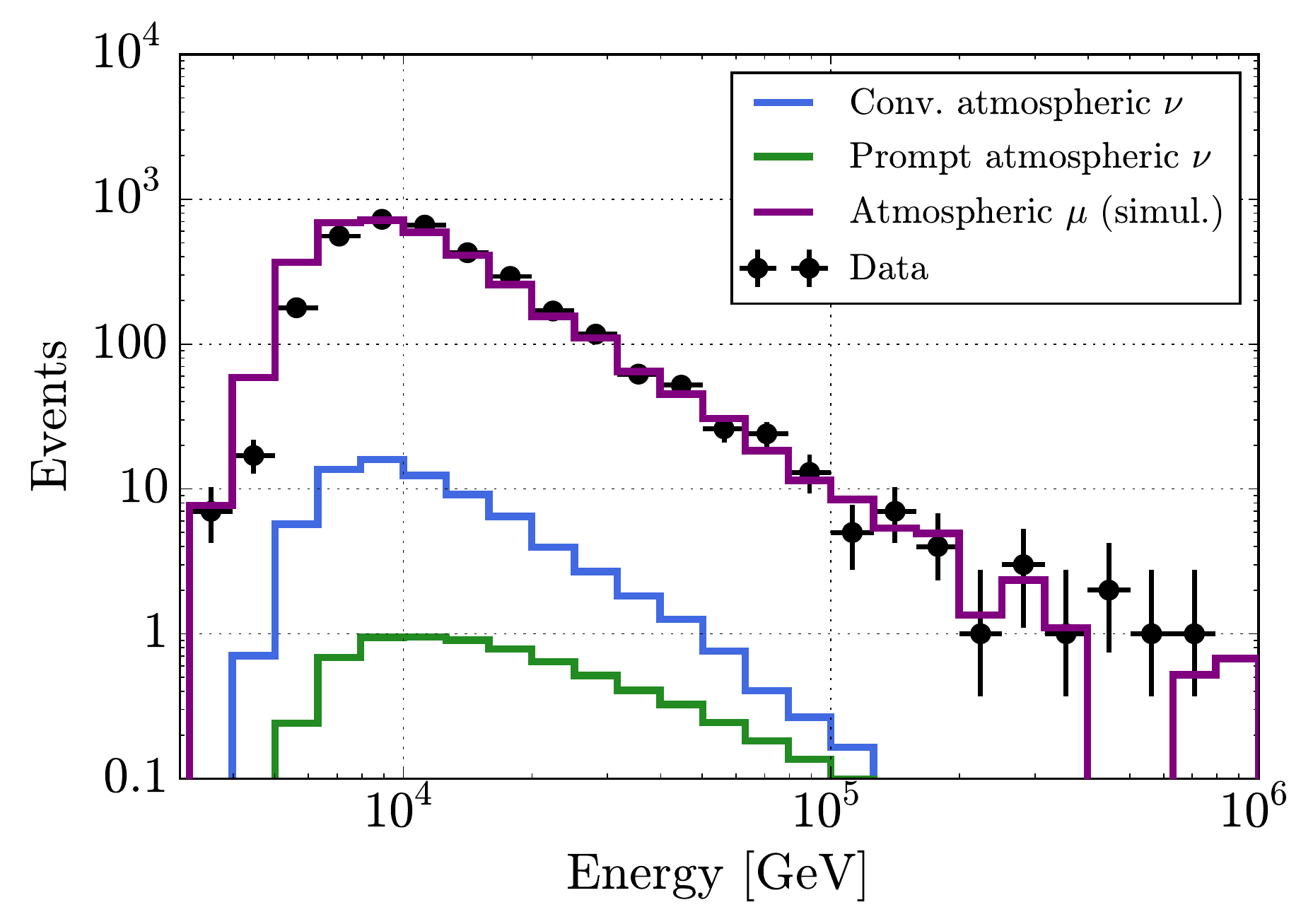}
  \caption{Distribution of deposited energy for the event sample of search S2 at next-to-final selection level.
  The lines show predictions for the conventional atmospheric neutrino flux (blue), the prompt atmospheric neutrino flux (green) and atmospheric muons (purple).
  The black data points represent experimental data, the error bars denote 68\% C.L. intervals as defined in \citet{feldman1998}.
  }
  \label{fig:appendix_edep_l4}
\end{figure}

The event sample of search S2 has been extended with respect to what was presented in \citet{schoenwald2013}.
Specifically, while only events with deposited energies larger than 38~TeV were analyzed in the original work, here we use all events with deposited energies above 7~TeV.
To convince ourselves that this is possible, we have carefully evaluated the behavior of the background components in the energy range that was added, in particular that of the residual atmospheric muon background.
Figure~\ref{fig:appendix_edep_l4} shows the distribution of deposited energies at next-to-final selection level, where the event sample is still dominated by atmospheric muons.
Above 7~TeV, our simulation matches the observed data well.
Below that energy, the selection efficiency rapidly drops and we observe a disagreement between simulation and experimental data.
We conclude that the simulation data below 7~TeV should not be used for analysis.

\begin{figure}
  \centering
  \includegraphics[width=.45\textwidth]{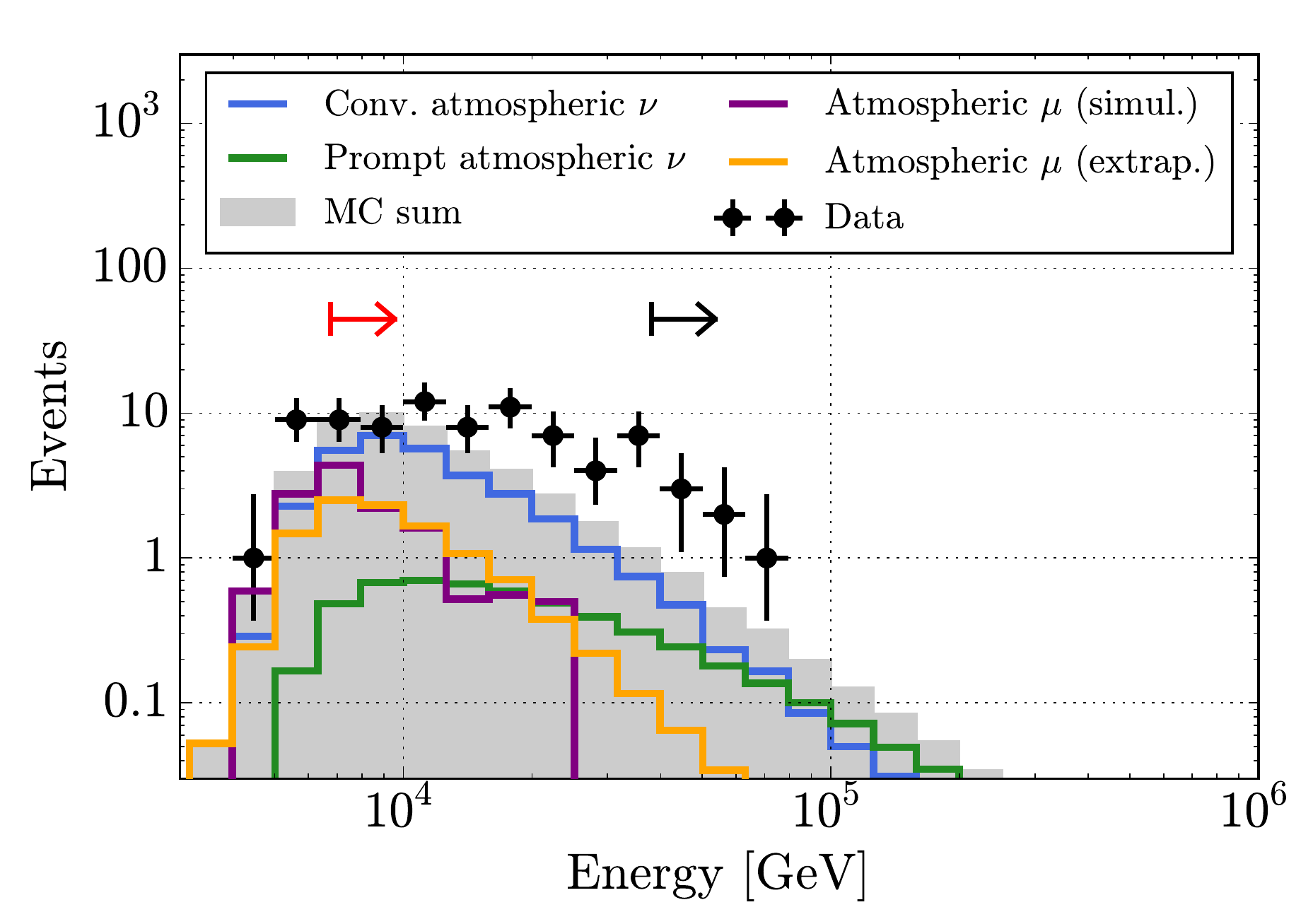}
  \caption{Same as Figure~\ref{fig:appendix_edep_l4}, but at final selection level.
  Additionally, an extrapolation of the muon background is shown in yellow.
  The gray filled histogram shows the sum of the two atmospheric neutrino components and the extrapolated muon background.
  The original work used events above 38~TeV (black arrow), here we use the data above 7~TeV (red arrow).
  }
  \label{fig:appendix_edep_l5}
\end{figure}

The distribution of deposited energies at final selection level is shown in Figure~\ref{fig:appendix_edep_l5}, the data above 7~TeV is used in this work.
Note that unlike in Figure~\ref{fig:data_all}, the simulation data has not been fitted to the experimental data for this Figure, but shows baseline predictions.
As before, predictions for atmospheric neutrinos are from \citet{honda2007} (conventional) and \citet{enberg2008} (prompt), both updated to the cosmic-ray composition model by \citet{gaisser2012} and corrected for the self-veto effect using the calculation from \citet{gaisser2014}.
The residual muon background has been determined with a CORSIKA simulation \citep{heck1998}, the simulated events remaining in the final sample are indicated by the purple histogram.
As this simulation data lacks statistics in particular at high energies, we have parametrized and extrapolated the distribution (yellow line).
This extrapolation is used as a pdf in the combined analysis presented here.
We note that at final level, the residual muon background is never the dominant component in the entire energy range and conclude that it is possible to use the data in the range between 7 and 38~TeV.

Figure~\ref{fig:appendix_edep_l5} also shows that even though the analysis in \citet{schoenwald2013} did not find an astrophysical component in the neutrino flux, there is an excess of experimental data above the baseline background predictions.
This excess was attributed to a very large prompt component in the original work, although with very large uncertainties.
We qualitatively reproduce this result in our individual fit of sample S2 (see Figure~\ref{fig:indiv_fits}, fourth row); we attribute the quantitative differences to the different energy range analyzed and the fact that the self-veto effect was neglected in the original work.
Furthermore, we observe that the excess is also compatible with an astrophysical flux as measured in the combined analysis presented here (cf. Figure~\ref{fig:data_all}, panel S2).
This shows that when analyzed individually, the data of sample S2 cannot be used to distinguish between a large prompt flux and a steep astrophysical flux.


\newpage
\begin{thebibliography}{}
\bibitem[Aartsen~et~al.(2013)]{hese2yearpaper} Aartsen, M. G., Abbasi, R., Abdou, Y., et al.\\(IceCube Collaboration) 2013, \href{http://dx.doi.org/10.1126/science.1242856}{Sci, 342, 1242856}
\bibitem[Aartsen~et~al.(2014a)]{moonshadowpaper} Aartsen, M. G., Abbasi, R., Abdou, Y., et al.\\(IceCube Collaboration) 2014a, \href{http://dx.doi.org/10.1103/PhysRevD.89.102004}{PhRvD, 89, 102004}
\bibitem[Aartsen~et~al.(2014b)]{energyrecopaper} Aartsen, M. G., Abbasi, R., Ackermann, M., et al.\\(IceCube Collaboration) 2014b, \href{http://dx.doi.org/10.1088/1748-0221/9/03/P03009}{JInst, 9, P03009}
\bibitem[Aartsen~et~al.(2014c)]{ic59numupaper} Aartsen, M. G., Abbasi, R., Ackermann, M., et al.\\(IceCube Collaboration) 2014c, \href{http://dx.doi.org/10.1103/PhysRevD.89.062007}{PhRvD, 89, 062007}
\bibitem[Aartsen~et~al.(2014d)]{ic40cascpaper} Aartsen, M. G., Abbasi, R., Ackermann, M., et al.\\(IceCube Collaboration) 2014d, \href{http://dx.doi.org/10.1103/PhysRevD.89.102001}{PhRvD, 89, 102001}
\bibitem[Aartsen~et~al.(2015a)]{ic7986numupaper} Aartsen, M. G., Abraham, K., Ackermann, M., et al.\\(IceCube Collaboration) 2015a, PhRvL, submitted, arXiv:\href{http://arxiv.org/abs/1507.04005}{1507.04005}
\bibitem[Aartsen~et~al.(2014e)]{hese3yearpaper} Aartsen, M. G., Ackermann, M., Adams, J., et al.\\(IceCube Collaboration) 2014e, \href{http://dx.doi.org/10.1103/PhysRevLett.113.101101}{PhRvL, 113, 101101}
\bibitem[Aartsen~et~al.(2014f)]{4yearpspaper} Aartsen, M. G., Ackermann, M., Adams, J., et al.\\(IceCube Collaboration) 2014f, \href{http://dx.doi.org/10.1088/0004-637X/796/2/109}{ApJ, 796, 109}
\bibitem[Aartsen~et~al.(2015b)]{ic7986hybridpaper} Aartsen, M. G., Ackermann, M., Adams, J., et al.\\(IceCube Collaboration) 2015b, \href{http://dx.doi.org/10.1103/PhysRevD.91.022001}{PhRvD, 91, 022001}
\bibitem[Aartsen~et~al.(2015c)]{flavorpaper} Aartsen, M. G., Ackermann, M., Adams, J., et al.\\(IceCube Collaboration) 2015c, \href{http://dx.doi.org/10.1103/PhysRevLett.114.171102}{PhRvL, 114, 171201}
\bibitem[Abbasi~et~al.(2010)]{pmtpaper} Abbasi, R., Abdou, Y., Abu-Zayyad, T., et al.\\(IceCube Collaboration) 2010, \href{http://dx.doi.org/10.1016/j.nima.2010.03.102}{NIMPA, 618, 139}
\bibitem[Abbasi~et~al.(2011)]{ic22cascpaper} Abbasi, R., Abdou, Y., Abu-Zayyad, T., et al.\\(IceCube Collaboration) 2011, \href{http://dx.doi.org/10.1103/PhysRevD.84.072001}{PhRvD, 84, 072001}
\bibitem[Abbasi~et~al.(2012)]{deepcorepaper} Abbasi, R., Abdou, Y., Abu-Zayyad, T., et al.\\(IceCube Collaboration) 2012, \href{http://dx.doi.org/10.1016/j.astropartphys.2012.01.004}{APh, 35, 615}
\bibitem[Abbasi~et~al.(2013)]{icetoppaper} Abbasi, R., Abdou, Y., Ackermann, M., et al.\\(IceCube Collaboration) 2013, \href{http://dx.doi.org/10.1016/j.nima.2012.10.067}{NIMPA, 700, 188}
\bibitem[Abbasi~et~al.(2009)]{daqpaper} Abbasi, R., Ackermann, M., Adams, J., et al.\\(IceCube Collaboration) 2009, \href{http://dx.doi.org/10.1016/j.nima.2009.01.001}{NIMPA, 601, 294}
\bibitem[Achterberg~et~al.(2006)]{firstyearperformancepaper} Achterberg, A., Ackermann, M., Adams, J., et al.\\(IceCube Collaboration) 2006, \href{http://dx.doi.org/10.1016/j.astropartphys.2006.06.007}{APh, 26, 155}
\bibitem[Achterberg~et~al.(2007)]{amandanumupaper} Achterberg, A., Ackermann, M., Adams, J., et al.\\(IceCube Collaboration) 2007, \href{http://dx.doi.org/10.1103/PhysRevD.76.042008}{PhRvD, 76, 042008}
\bibitem[Ackermann~et~al.(2015)]{fermidiffuse} Ackermann, M., Ajello, M., Albert, A., et al.\\(Fermi LAT Collaboration) 2015, \href{http://dx.doi.org/10.1088/0004-637X/799/1/86}{ApJ, 799, 86}
\bibitem[Adri\'an-Mart\'inez~et~al.(2014)]{antarespspaper} Adri\'an-Mart\'inez, S., Albert, A., Andr\'e, M., et al.\\(ANTARES Collaboration) 2014, \href{http://dx.doi.org/10.1088/2041-8205/786/1/L5}{ApJL, 786, L5}
\bibitem[Athar~et~al.(2006)]{athar2006} Athar, H., Kim, C. S., \& Lee, J. 2006, \href{http://dx.doi.org/10.1142/S021773230602038X}{MPLA, 21, 1049}
\bibitem[Beacom~et~al.(2003)]{beacom2003} Beacom, J. F., Bell, N. F., Hooper, D., Pakvasa, S.,\\ \& Weiler, T. J. 2003, \href{http://dx.doi.org/10.1103/PhysRevD.68.093005}{PhRvD, 68, 093005}
\bibitem[Becker(2008)]{becker2008} Becker, J. K. 2008, \href{http://dx.doi.org/10.1016/j.physrep.2007.10.006}{PhR, 458, 173}
\bibitem[Bednarek~et~al.(2005)]{bednarek2005} Bednarek, W., Burgio, G. F., \& Montaruli, T. 2005, \href{http://dx.doi.org/10.1016/j.newar.2004.11.001}{NewAR, 49, 1}
\bibitem[Bhattacharya~et~al.(2015)]{bhattacharya2015} Bhattacharya, A., Enberg, R., Reno, M. H., Sarcevic, I.,\\ \& Stasto, A. 2015, \href{http://dx.doi.org/10.1007/JHEP06(2015)110}{JHEP, 06, 110}
\bibitem[Bugaev~et~al.(1989)]{bugaev1989} Bugaev, E. V., Naumov, V. A., Sinegovsky, S. I.,\\ \& Zaslavskaya, E. S. 1989, \href{http://dx.doi.org/10.1007/BF02509070}{NCimC, 12, 41}
\bibitem[Bustamante~et~al.(2015)]{bustamante2015} Bustamante, M., Beacom, J. F., \& Winter, W. 2015, arXiv:\href{http://arxiv.org/abs/1506.02645}{1506.02645}
\bibitem[Choubey~\&~Rodejohann(2009)]{choubey2009} Choubey, S., \& Rodejohann, W. 2009, \href{http://dx.doi.org/10.1103/PhysRevD.80.113006}{PhRvD, 80, 113006}
\bibitem[Enberg~et~al.(2008)]{enberg2008} Enberg, R., Reno, M. H., \& Sarcevic, I. 2008, \href{http://dx.doi.org/10.1103/PhysRevD.78.043005}{PhRvD, 78, 043005}
\bibitem[Feldman~\&~Cousins(1998)]{feldman1998} Feldman, G. J., \& Cousins, R. D. 1998, \href{http://dx.doi.org/10.1103/PhysRevD.57.3873}{PhRvD, 57, 3873}
\bibitem[Gaggero~et~al.(2015)]{gaggero2015} Gaggero, D., Grasso, D., Marinelli, A., Urbano, A.,\\ \& Valli, M. 2015, arXiv:\href{http://arxiv.org/abs/1504.00227}{1504.00227}
\bibitem[Gaisser(1990)]{gaisser1990} Gaisser, T. K. 1990, Cosmic Rays and Particle Physics (Cambridge University Press)
\bibitem[Gaisser~et~al.(1995)]{gaisser1995} Gaisser, T. K., Halzen, F., \& Stanev, T. 1995, \href{http://dx.doi.org/10.1016/0370-1573(95)00003-Y}{PhR, 258, 173}
\bibitem[Gaisser(2012)]{gaisser2012} Gaisser, T. K. 2012, \href{http://dx.doi.org/10.1016/j.astropartphys.2012.02.010}{APh, 35, 801}
\bibitem[Gaisser~et~al.(2014)]{gaisser2014} Gaisser, T. K., Jero, K., Karle, A., \& van Santen, J. 2014, \href{http://dx.doi.org/10.1103/PhysRevD.90.023009}{PhRvD, 90, 023009}
\bibitem[Gandhi~et~al.(1996)]{gandhi1996} Gandhi, R., Quigg, C., Reno, M. H., \& Sarcevic, I. 1996,\\ \href{http://dx.doi.org/10.1016/0927-6505(96)00008-4}{APh, 5, 81}
\bibitem[Gonzalez-Garcia~et~al.(2014)]{gonzalezgarcia2014} Gonzalez-Garcia, M. C., Maltoni, M., \& Schwetz, T. 2014,\\ \href{http://dx.doi.org/10.1007/JHEP11(2014)052}{JHEP, 11, 052}
\bibitem[Guetta~et~al.(2004)]{guetta2004} Guetta, D., Hooper, D., Alvarez-Mu\~{n}iz, J., Halzen, F.,\\ \& Reuveni, E. 2004, \href{http://dx.doi.org/10.1016/S0927-6505(03)00211-1}{APh, 20, 429}
\bibitem[Heck~et~al.(1998)]{heck1998} Heck, D., Knapp, J., Capdevielle, J. N., Schatz, G., \& Thouw, T. 1998, Tech. Rep. FZKA 6019, Forschungszentrum Karlsruhe
\bibitem[Honda~et~al.(2007)]{honda2007} Honda, M., Kajita, T., Kasahara, K., Midorikawa, S.,\\ \& Sanuki, T. 2007, \href{http://dx.doi.org/10.1103/PhysRevD.75.043006}{PhRvD, 75, 043006}
\bibitem[Hooper~et~al.(2003)]{hooper2003} Hooper, D., Nunokawa, H., Peres, O. L. G.,\\ \& Zukanovich Funchal, R. 2003, \href{http://dx.doi.org/10.1103/PhysRevD.67.013001}{PhRvD, 67, 013001}
\bibitem[Kappes~et~al.(2007)]{kappes2007} Kappes, A., Hinton, J., Stegmann, C., \& Aharonian, F. 2007, \href{http://dx.doi.org/10.1086/508936}{ApJ, 656, 870}
\bibitem[Kashti~\&~Waxman(2005)]{kashti2005} Kashti, T., \& Waxman, E. 2005, \href{http://dx.doi.org/10.1103/PhysRevLett.95.181101}{PhRvL, 95, 181101}
\bibitem[Kistler~\&~Beacom(2006)]{kistler2006} Kistler, M. D., \& Beacom, J. F. 2006, \href{http://dx.doi.org/10.1103/PhysRevD.74.063007}{PhRvD, 74, 063007}
\bibitem[Klein~et~al.(2013)]{klein2013} Klein, S. R., Mikkelsen, R. E., \& Becker Tjus, J. 2013,\\ \href{http://dx.doi.org/10.1088/0004-637X/779/2/106}{ApJ, 779, 106}
\bibitem[Laha~et~al.(2013)]{laha2013} Laha, R., Beacom, J. F., Dasgupta, B., Horiuchi, S.,\\ \& Murase, K. 2013, \href{http://dx.doi.org/10.1103/PhysRevD.88.043009}{PhRvD, 88, 043009}
\bibitem[Learned~\&~Mannheim(2000)]{learned2000} Learned, J. G., \& Mannheim, K. 2000, \href{http://dx.doi.org/10.1146/annurev.nucl.50.1.679}{ARNPS, 50, 679}
\bibitem[Learned~\&~Pakvasa(1995)]{learned1995} Learned, J. G., \& Pakvasa, S. 1995, \href{http://dx.doi.org/10.1016/0927-6505(94)00043-3}{APh, 3, 267}
\bibitem[Lipari~et~al.(2007)]{lipari2007} Lipari, P., Lusignoli, M., \& Meloni, D. 2007, \href{http://dx.doi.org/10.1103/PhysRevD.75.123005}{PhRvD, 75, 123005}
\bibitem[Loeb~\&~Waxman(2006)]{loeb2006} Loeb, A., \& Waxman, E. 2006, \href{http://dx.doi.org/10.1088/1475-7516/2006/05/003}{JCAP, 05, 003}
\bibitem[Martin~et~al.(2003)]{martin2003} Martin, A. D., Ryskin, M. G., \& Stasto, A. M. 2003,\\ \href{http://www.actaphys.uj.edu.pl/vol34/abs/v34p3273.htm}{AcPPB, 34, 3273}
\bibitem[Mena~et~al.(2014)]{mena2014} Mena, O., Palomares-Ruiz, S., \& Vincent, A. C. 2014,\\ \href{http://dx.doi.org/10.1103/PhysRevLett.113.091103}{PhRvL, 113, 091103}
\bibitem[M\"ucke~et~al.(2003)]{muecke2003} M\"ucke, A., Protheroe, R. J., Engel, R., Rachen, J. P.,\\ \& Stanev, T. 2003, \href{http://dx.doi.org/10.1016/S0927-6505(02)00185-8}{APh, 18, 593}
\bibitem[Murase~et~al.(2013)]{murase2013} Murase, K., Ahlers, M., \& Lacki, B. C. 2013,\\ \href{http://dx.doi.org/10.1103/PhysRevD.88.121301}{PhRvD, 88, 121301(R)}
\bibitem[Murase~et~al.(2008)]{murase2008} Murase, K., Inoue, S., \& Nagataki, S. 2008, \href{http://dx.doi.org/10.1086/595882}{ApJL, 689, L105}
\bibitem[Neronov~\&~Semikoz(2014)]{neronov2014} Neronov, A., \& Semikoz, D. 2014, arXiv:\href{http://arxiv.org/abs/1412.1690}{1412.1690}
\bibitem[Olive~et~al.(2014)]{olive2014} Olive, K. A., Agashe, K., Amsler, C., et al. (Particle Data Group) 2014, \href{http://dx.doi.org/10.1088/1674-1137/38/9/090001}{ChPhC, 38, 090001}
\bibitem[Pagliaroli~et~al.(2015)]{pagliaroli2015} Pagliaroli, G., Palladino, A., Vissani, F., \& Villante, F. L. 2015, arXiv:\href{http://arxiv.org/abs/1506.02624}{1506.02624}
\bibitem[Palladino~et~al.(2015)]{palladino2015} Palladino, A., Pagliaroli, G., Villante, F. L., \& Vissani, F. 2015, \href{http://dx.doi.org/10.1103/PhysRevLett.114.171101}{PhRvL, 114, 171101}
\bibitem[Palomares-Ruiz~et~al.(2015)]{palomaresruiz2015} Palomares-Ruiz, S., Vincent, A. C., \& Mena, O. 2015,\\ \href{http://dx.doi.org/10.1103/PhysRevD.91.103008}{PhRvD, 91, 103008}
\bibitem[Sch\"onert~et~al.(2009)]{schoenert2009} Sch\"onert, S., Gaisser, T. K., Resconi, E., \& Schulz, O. 2009, \href{http://dx.doi.org/10.1103/PhysRevD.79.043009}{PhRvD, 79, 043009}
\bibitem[Sch\"onwald~et~al.(2013)]{schoenwald2013} Sch\"onwald, A., Brown, A., \& Mohrmann, L. for the IceCube Collaboration 2013, in Proc. 33rd Int. Cosmic Ray Conf., 0662 (arXiv:\href{http://arxiv.org/abs/1309.7003}{1309.7003})
\bibitem[Senno~et~al.(2015)]{senno2015} Senno, N., M\'esz\'aros, P., Murase, K., Baerwald, P.,\\ \& Rees, M. J. 2015, \href{http://dx.doi.org/10.1088/0004-637X/806/1/24}{ApJ, 806, 24}
\bibitem[Spiering(2012)]{spiering2012} Spiering, C. 2012, \href{http://dx.doi.org/10.1140/epjh/e2012-30014-2}{EPJH, 37, 515}
\bibitem[Stecker~et~al.(1991)]{stecker1991} Stecker, F. W., Done, C., Salamon, M. H., \& Sommers, P. 1991, \href{http://dx.doi.org/10.1103/PhysRevLett.66.2697}{PhRvL, 66, 2697}
\bibitem[Vissani~et~al.(2013)]{vissani2013} Vissani, F., Pagliaroli, G., \& Villante, F. L. 2013, \href{http://dx.doi.org/10.1088/1475-7516/2013/09/017}{JCAP, 09, 017}
\bibitem[Waxman~\&~Bahcall(1997)]{waxman1997} Waxman, E., \& Bahcall, J. 1997, \href{http://dx.doi.org/10.1103/PhysRevLett.78.2292}{PhRvL, 78, 2292}
\bibitem[Wilks(1938)]{wilks1938} Wilks, S. S. 1938, Ann. Math. Stat., 9, 60
\bibitem[Winter(2013)]{winter2013} Winter, W. 2013, \href{http://dx.doi.org/10.1103/PhysRevD.88.083007}{PhRvD, 88, 083007}
\end{thebibliography}
\end{document}